\documentclass{pasa}

\usepackage{graphicx}

\title[Atmospheres of Exoplanets and Brown Dwarfs]{The Dawes Review 3: The Atmospheres of Extrasolar Planets and Brown Dwarfs}
\author[Bailey]{Jeremy Bailey$^1$\\
\affil{$^1$School of Physics, University of New South Wales, NSW, 2052,
Australia}}%
\jid{PASA}
\doi{10.1017/pasa.\the\year.xxx}
\jyear{\the\year}

\usepackage[authoryear]{natbib}
\bibpunct{(}{)}{;}{a}{}{,}
\def\earth{\hbox{$\oplus$}}

\begin{document} 
\begin{abstract} 
The last few years has seen a dramatic increase in the number of exoplanets known and in
the range of methods for characterising their atmospheric properties. At the same time, new discoveries of increasingly cooler
brown dwarfs have pushed down their temperature range which now extends down to Y-dwarfs of $<$300 K. Modelling of these
atmospheres has required the development of new techniques to deal with the molecular chemistry and clouds in these objects.
The atmospheres of brown dwarfs are relatively well understood, but some problems remain, in particular the behavior of clouds
at the L/T transition. Observational data for exoplanet atmosphere characterization is largely limited to giant exoplanets that
are hot because they are near to their star (hot Jupiters) or because they are young and still cooling. For these planets there
is good evidence for the presence of CO and H$_2$O absorptions in the IR. Sodium absorption is observed in a number of objects.
Reflected light measurements show that some giant exoplanets are very dark, indicating a cloud free atmosphere. However, there
is also good evidence for clouds and haze in some other planets. It is also well established that some highly irradiated
planets have inflated radii, though the mechanism for this inflation is not yet clear. Some other issues in the composition and
structure of giant exoplanet atmospheres such as the occurence of inverted temperature structures, the presence or absence of
CO$_2$ and CH$_4$, and the occurrence of high C/O ratios are still the subject of investigation and debate.  
\end{abstract}


\maketitle 

{\it The Dawes Reviews are substantial reviews of topical areas in astronomy, published by authors of international standing at the
invitation of the PASA Editorial Board. The reviews recognise William Dawes (1762--1836), second lieutenant in the Royal
Marines and the astronomer on the First Fleet. Dawes was not only an accomplished astronomer, but spoke five languages, had a
keen interest in botany, mineralogy, engineering, cartography and music, compiled the first Aboriginal-English disctionary, and
was an outspoken opponent of slavery.}

\section{INTRODUCTION } \label{sec:intro}

It is appropriate to consider the properties of extrasolar planet and brown dwarf
atmospheres together because they have many similarities. Planets and brown dwarfs cover similar temperature
ranges and have similar radii. Planets extend up from very low temperatures (such as those of the ice giants Uranus
and Neptune in our Solar system) to effective temperatures of $\sim$3000 K in hot Jupiters, while new
discoveries are continually pushing down the temperature of the coolest known brown dwarfs. 
The recently discovered Y dwarf class have temperatures as low as $\sim$300 K \citep{cushing11}.

The important processes that occur in these atmospheres are also similar as
these are determined primarily by effective temperature. Molecules, chemistry and clouds are important
in determining the opacities and hence structure of all these objects. At any temperature
below about $\sim$2000K, solid and liquid condensates can start to form, resulting in considerable
complications compared with higher temperatures where only gas phase processes need to be
considered. More complex molecules such as methane (CH$_4$) become important and the excitation
of high vibrational and rotational levels mean that vast numbers
of spectral lines are needed to model the opacity. The modelling of these atmospheres thus presents new
challenges compared with those encountered in conventional stellar atmosphere models, and
these challenges are largely common to the modeling of both exoplanets and brown
dwarfs. The differences between giant exoplanets and brown dwarfs include the generally lower mass (and hence gravity) in exoplanets, and the
difference in environment. An exoplanet orbits a star, and the stellar illumination can have
a significant influence on its structure and properties, particularly for close in planets such as hot Jupiters. 
The presence of the host star also impacts on our ability to observe the planet. While some observations 
can be easier for planets than brown dwarfs (e.g.
determining mass and
radius), spectroscopy to characterise the atmospheres is usually extremely challenging for exoplanets while 
relatively straightforward for brown dwarfs.

The structure of this
review will be to begin with looking at brown dwarf atmospheres. This reflects the fact that
these are better observed and understood objects, without the complications that are
introduced by the presence of the host star in exoplanet systems, but nevertheless
illustrate many of the processes that are also important in giant exoplanets. 
Exoplanet atmospheres will be considered next, with a brief look at the giant planets in our
own solar system as a guide. Observations relevant to atmospheric structure and composition
are now being obtained by a number of methods primarily for giant exoplanets. These
will be outlined and the results of these methods discussed.

The next section will look at the modelling of brown dwarf and exoplanet atmospheres. The techniques are very
similar for both classes of objects. A final section will look at the possibilities of detecting extrasolar
habitable planets and searching for signatures of life on such planets.

\section{BROWN DWARFS}

\subsection{History and Properties}

The existence of brown dwarfs was predicted long before they were recognized
observationally. \citet{kumar63} and \citet{hayashi63} showed that there was a lower limit to the mass of a
main-sequence star below which hydrogen burning could not occur. Kumar referred to the objects
below this limit as ``black'' dwarfs, but the name ``brown dwarf'' proposed by 
\citet{tarter75} is the one that has been adopted. More recent models set the hydrogen
burning mass limit at 0.072--0.075 M$_{\odot}$ for a solar composition and somewhat higher
for lower metallicities \citep{chabrier97,burrows01}.

\begin{figure}
\begin{center}
\includegraphics[scale=0.44, angle=0]{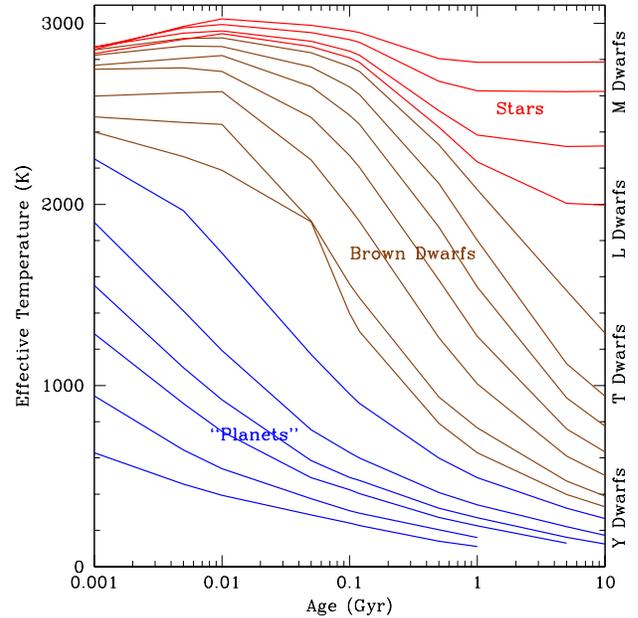}
\caption{Evolution of effective temperature for objects from 0.0005 to 0.1 M$_{\odot}$ based on the models
of \citet{baraffe03}. The red tracks are for stars with masses above the hydrogen burning
limit. The magenta tracks are for brown dwarfs, and the blue tracks are for objects below
the deuterium burning limit (planets or sub brown dwarfs) The tracks plotted from top to bottom are masses of (Stars: 0.1,
0.09, 0.08, 0.075 M$_{\odot}$) (Brown Dwarfs: 0.07, 0.06, 0.05, 0.04, 0.03, 0.02, 0.015 
M$_{\odot}$), (Planets: 0.01, 0.005, 0.003, 0.002, 0.001, 0.0005 M$_{\odot}$).} \label{fig1}
\end{center}
\end{figure}

Because brown dwarfs do not have a continuing nuclear energy source, their evolution is a process of
cooling and decreasing luminosity with age. Unlike stars on the main sequence whose properties are
determined primarily by their mass, the temperature and luminosity of a brown dwarf are
determined by both mass and age. This is illustrated in figure \ref{fig1} where the
effective temperatures of low mass objects with masses between 0.0005 and 0.1 M$_{\odot}$ are
plotted as a function of age. The
evolutionary models used here are those of \citet{baraffe03} but similar general
trends would be obtained with other recent models \citep[e.g][]{chabrier00a,
baraffe02,burrows01,saumon08}. 

The four tracks at the top of figure \ref{fig1} are for objects massive enough to be stars, so their
effective temperature eventually stabilizes to a near constant value. However, for brown
dwarfs the effective temperature continues to decrease with increasing age. It can be seen
from this diagram that a determination of effective temperature alone is not sufficient to
determine whether an object is a star or a brown dwarf. An object with T$_{eff}$ = 2200 K,
for example, can be a young brown dwarf or an older star.

This age-mass degeneracy complicated the early observational search for brown dwarfs, and
while several candidates were found  \citep[e.g. GD165b][]{becklin88} it was not possible
to confirm them as brown dwarfs. That changed in 1995 with the discovery of Gl 229b
\citep{nakajima95,oppenheimer95}, an object sufficiently cool to be unambiguosly a brown
dwarf, and with the use of the lithium test to confirm the brown dwarf nature of several
objects in the Pleiades cluster \citep{rebolo95,rebolo96,basri96}. The lithium test
\citep{rebolo92} relies on the fact that lithium is destroyed by nuclear reactions down to
masses just below the hydrogen burning limit. Since cool dwarfs are fully convective, lithium
would be removed from the photosphere if these reactions occurred. Hence the presence of lithium in the spectrum
can be used to confirm that an object is a brown dwarf.

The deuterium burning mass limit which is at about 13 M$_J$ or 0.0125 M$_{\odot}$ \citep{saumon96,burrows97,chabrier00b} is
usually considered to be the lower mass limit for brown dwarfs. Objects below this mass limit that orbit stars are generally
agreed to be designated as planets. There is less consensus on how to refer to object below this mass limit that do not orbit a
star. While these are sometimes referred to as ``free-floating planets'' \citep{lucas00,delorme12} it has also been argued that
such objects should not be referred to as planets but as ``sub-brown dwarfs'' or some other designation \citep[see][for a
discussion of the issues involved in this controversy]{boss03,basri06}.

\begin{figure}
\begin{center}
\includegraphics[scale=0.44, angle=0]{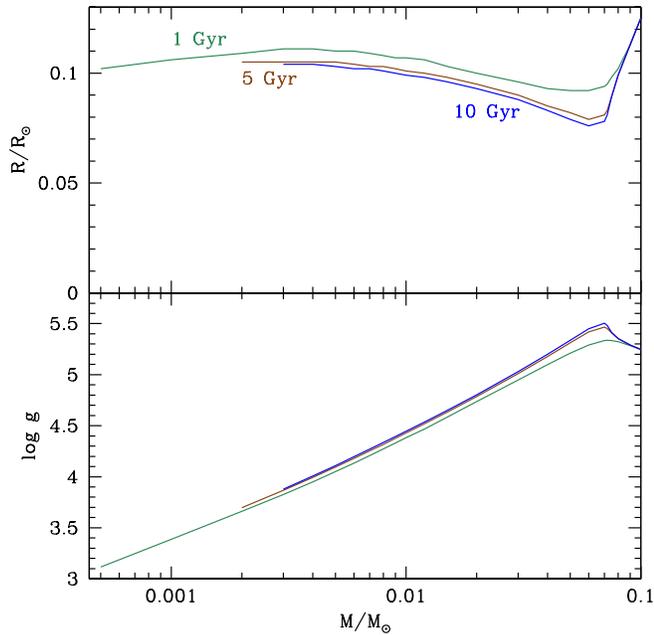}
\caption{Radius and surface gravity (log g in cgs units) as function of mass for the models of \citet{baraffe03} at ages of 1 Gyr, 5 GYr and 10 Gyr.} \label{fig_rad}
\end{center}
\end{figure}

The electron degeneracy in the cores of brown dwarfs results in their radius varying little with mass as can be seen in figure \ref{fig_rad}. All
brown dwarfs (except at very young ages) have radii not far from 0.1 R$_\odot$ or about 1 Jupiter radius. A consequence of this
is that surface gravity ($g = GM/R^2$) varies with mass from more than 1000 m s$^{-2}$ (log g = 5 in cgs units) to around 30 m
s$^{-2}$ for Jupiter mass objects as shown in the lower panel of figure \ref{fig_rad}.

\begin{figure}
\begin{center}
\includegraphics[scale=0.44, angle=0]{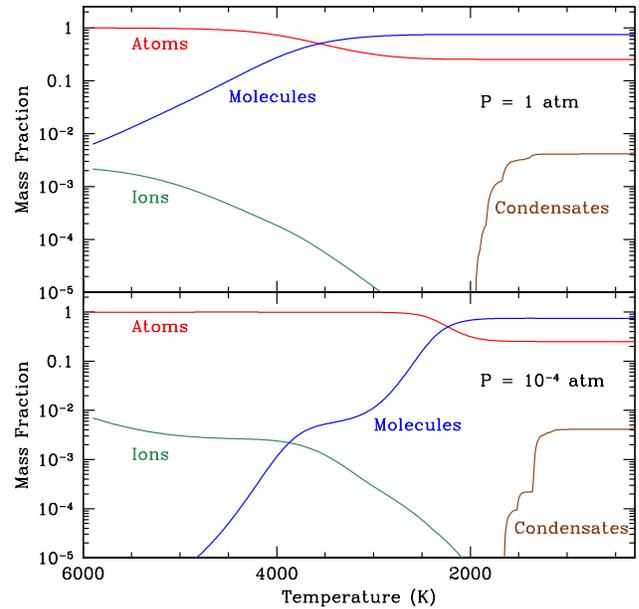}
\caption{Equilibrium composition of a gas with solar elemental abundances as a function of temperature at two different pressures using the chemical model of \citet{bailey12}.} \label{fig2}
\end{center}
\end{figure}

Brown dwarfs are objects whose atmospheric composition is dominated by molecular gas, as opposed to atoms and ions in the case of hotter stars. This is apparent from figure \ref{fig2} which shows the chemical equilibrium composition of a solar composition gas \citep*[using the abundances of][]{grevesse07}. It shows the division of the material by mass fraction into ions, atoms, gas-phase molecules and solid or liquid condensates as calculated by the chemical model of \cite{bailey12}.
It can be seen that molecules become dominant over atoms for temperatures below about 3500 K. Helium and other noble gases persist as atoms at all temperatures, but other elements are mostly in the form of molecules. Below about 2000 K condensed phases start to appear, and become a significant fraction of the material. At lower pressures, as shown in the lower panel, the pattern is similar but shifted to lower temperatures.

The number of ultracool dwarfs\footnote{ultracool dwarf is a name normally used for objects with spectral type later than
about M7, which could potentially be brown dwarfs, but could also be stars} has increased rapidly over the years since
the recognition of the first brown dwarfs in 1995. Most of the objects have come from deep surveys such as the Sloan
Digital Sky Survey \citep[SDSS ---][]{fan00,hawley02} and the Canada-France Brown Dwarfs Survey \citep[CFBDS ---
][]{delorme08a, albert11} and particularly from infrared surveys such as the Deep Near-Infrared Sky Survey \citep[DENIS
---][]{delfosse97,martin04}, the 2 Micron All Sky Survey \citep[2MASS ---][]{kirkpatrick00,burgasser02,burgasser04}, and
the UKIRT Infrared Deep Sky Survey \citep[UKIDSS ---][]{pinfield08,burningham10,burningham13}.

The most recent additions have come from the Wide-field Infrared Survey Explorer \citep[WISE ---][]{wright10}. This Earth
orbiting NASA mission surveyed the entire sky at four wavelengths (3.4, 4.6, 12 and 22 $\mu$m). The first of these
wavelengths probes a deep CH$_4$ absorption band in brown dwarfs. WISE has proved effective in identifying the coolest
brown dwarfs. It has led to the discovery of many T dwarfs \citep{kirkpatrick11,mace13} and to the first Y dwarfs
\citep{cushing11,kirkpatrick12,tinney12}.

Other recent discoveries from WISE are that of a binary brown dwarf \citep{luhman13} and an extremely cool brown dwarf
\citep{luhman14} both at distances of around 2pc. WISE
J104915.57$-$531906.1 (also known as Luhman 16) consists of an L7.5 -- L8 primary and T0.5 -- T1.5 
secondary \citep{burgasser13,kniazev13}. Its brightness and proximity are likely to make it an important object for future
detailed studies. WISE J088510.83$-$071442.5 \citep{luhman14} appears to be the coldest brown dwarf known based on its absolute
magnitude and colours. These two systems
are the closest brown dwarf systems, and the third and fourth closest systems to the Earth (after the $\alpha$
Centauri system and Barnard's star).

\subsection{Brown Dwarf Spectral Sequence}

The study of brown dwarfs has led to a significant extension of the traditional spectral sequence from O-M that was
adopted more than 100 years ago \citep{cannon01}. Objects such as GD 165B and Gl 229B clearly had quite different spectra
and were cooler objects than any M dwarfs. This was recognized by the adoption of the new spectral classes L and T. The
motivation for this and the reasons for the choice of those letters are described by \citet{kirkpatrick99}. The sequence
has been further extended by the recent recognition of even cooler objects that have been assigned to the new spectral
class Y \citep{cushing11}.

\begin{figure*}
\begin{center}
\includegraphics[scale=0.85, angle=0]{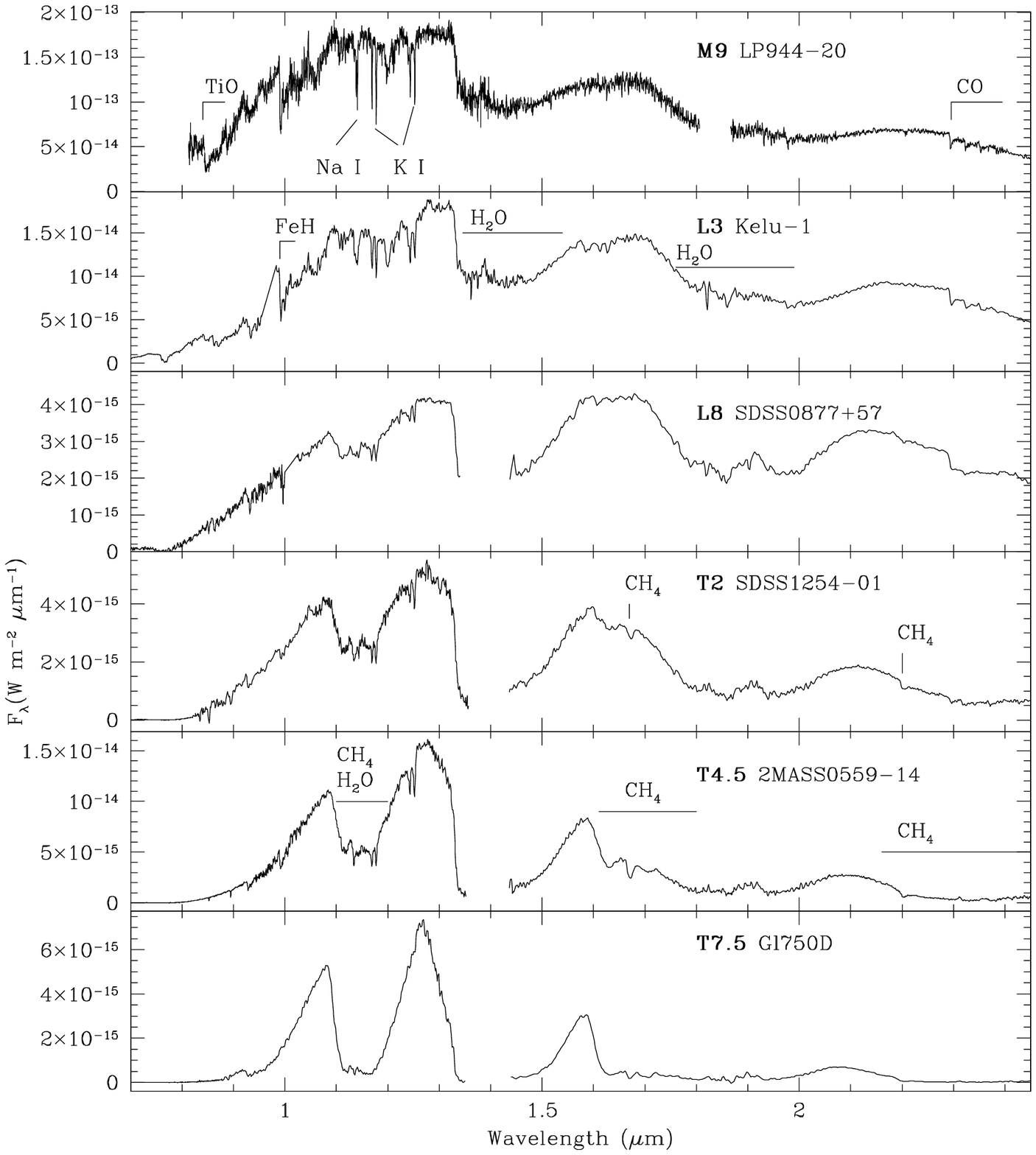}
\caption{Spectra of ultracool dwarfs from M9 to T7.5. The species responsible for the main absorption features are indicated. Spectral data is from \citet{burgasser03}, \citet{cushing05}, \citet{geballe01}, \citet{geballe02}, \citet{leggett00}, \citet{leggett01}, \citet{leggett02}, \citet{rayner09}, \citet{ruiz97}} \label{fig_spec}
\end{center}
\end{figure*}

Figure \ref{fig_spec} shows the main features of the spectral sequence from M9 to T7.5 with the main absorbing species
indicated.

\subsubsection{M Dwarfs}

The M spectral classification has been recognized from the early days of astronomical spectroscopy. While most M dwarfs
are stars, young objects of late M spectral types can be brown dwarfs (as shown in figure \ref{fig1}). The modern
classification scheme for M-dwarfs is based on that of \citet{boeshaar76} extended by \citet{boeshaar85} and
\citet*{kirkpatrick91} to spectral type M9.5. The \citet{kirkpatrick91} spectral classification is based on the spectral
region from 630-900 nm. The spectral standards chosen for late M types are listed in table \ref{tab_stand}.

The M spectral class is characterized by the presence of bands of TiO and VO. TiO bands increase in strength up to
spectral type M6, and VO becomes strong in the latest types.

In the near infrared (near-IR) M dwarfs show broad absorptions due to H$_2$O centered around 1.4 and 1.9 $\mu$m
increasing in strength with later spectral types. Late M dwarfs also show Na I and K I absorptions in the 1.15--1.25
$\mu$m region. FeH absorption is present in the Wing-Ford band at 1 $\mu$m as well as the E-A band in the 1.6 $\mu$m
region \citep{hargreaves10}. CO absorption is present at 2.3 $\mu$m.

\begin{table*}
\caption{Mean Properties and Spectral Standards for late M to Y dwarfs}\label{tab_stand}
\begin{center}
\begin{tabular}{lrlllll}
\hline\hline
  & T$_{\mbox{\scriptsize eff}}$ & M$_J$ & M$_H$ & M$_K$ & Spectral Standard & Refs \\
\hline M7 & 2633 & 10.31 &  9.94 &  9.50 & 2MASS J16553529$-$0823401 (Gl 644C,  vB8)& 1, 7 \\
M8 & 2520 & 10.99 & 10.40 &  9.88 & 2MASS J19165762+0509021 (Gl 752B, vB10) & 1, 7 \\
M9 & 2465 & 11.80 & 11.15 & 10.62 & 2MASS J14284323+3310391 (LHS 2924)      & 1, 7\\
L0 & 2438 & 11.69 & 11.04 & 10.46 & 2MASS J03454316+2540233                   & 2, 7 \\
L1 & 2329 & 11.87 & 11.26 & 10.66 & 2MASS J07464256+2000321                   & 2, 7 \\
L2 &      & 12.18 & 11.45 & 10.82 & 2MASS J13054019$-$2541059 (Kelu-1)        & 2, 7 \\
L3 & 1948 & 12.81 & 11.97 & 11.26 & 2MASS J11463449+2230527                   & 2, 7 \\
L4 & 1801 & 12.83 & 12.14 & 11.25 & 2MASS J11550087+2307058                   & 2, 7 \\
L5 & 1686 & 13.44 & 12.61 & 11.96 & 2MASS J12281523$-$1547342                 & 2, 7 \\
L6 & 1501 & 14.12 & 13.05 & 12.00 & 2MASS J08503593+1057156                   & 2, 7 \\
L7 & 1446 & 14.67 & 13.70 & 12.89 & 2MASS J02059240$-$1159296                 & 2, 7 \\
L8 & 1445 & 14.68 & 13.77 & 13.05 & 2MASS J16322911$-$1904407                 & 2, 7 \\
L9 &      & 14.33 & 13.48 & 12.73 & 2MASS J02550327$-$4700509                 & 3, 7 \\
T0 & 1370 & 14.24 & 13.52 & 13.17 & SDSS J120747.17+024424.8                  & 4, 7 \\
T1 &      & 14.37 & 13.81 & 13.62 & SDSS J083717.21$-$000018.0                & 4, 7 \\
T2 & 1328 & 14.43 & 13.88 & 13.58 & SDSS J125453.90$-$012247.4                & 4, 7 \\
T3 &      &       &       &       & 2MASS J12095613$-$1004008                 & 4, 7 \\
T4 & 1251 & 15.04 & 14.41 & 14.13 & 2MASS J22541892+3123498                   & 4, 7 \\
T5 & 1185 & 14.43 & 14.66 & 14.81 & 2MASS J15031961+2525196                   & 4, 7 \\
T6 & 1001 & 15.22 & 15.56 & 15.77 & SDSS J1642414.37+002915.6                 & 4, 7 \\
T7 &  820 & 15.54 & 15.97 & 16.01 & 2MASS J07271824+1710012                   & 4, 7 \\
T8 &  638 & 16.43 & 16.82 & 16.93 & 2MASS J04151954$-$0935066                 & 4, 8 \\
T9 &  565 & 18.39 & 18.77 & 18.89 & UGPS J072227.51$-$054031.2                & 5, 8 \\
Y0 &  371 & 20.09 & 20.60 & 20.70 & WISE J173835.52+273258.9                  &  5, 8 \\
Y1 &      & &  &  & WISE J035000.32$-$565830.2                  &  6 \\
\hline\hline
\end{tabular}
\end{center}
\medskip
{\bf References.} First reference is to adoption of the spectral standard, and the second reference is to the source of the mean absolute magitudes. \\
1. \citet{kirkpatrick91}, 2. \citet{kirkpatrick99}, 3. \citet{kirkpatrick10}
4. \citet{burgasser06a}, 5. \citet{cushing11}, 6. \citet{kirkpatrick12}, 7. \citet{dupuy12},
8. \citet{dupuy13} \\
{\bf Notes.}\\
Mean effective temperatures are from the data of figure \ref{fig_teff}.
Spectral standards are those adopted for optical classification up to spectral class L8, and for near-IR classification for L9 and later. Near-IR spectral standards for earlier types can be found in \citet{kirkpatrick10}. Spectral data is available for download for most of these objects (and other late-type dwarfs) at: \\
SpeX Prism Spectral Libraries (A. Burgasser) \\
--- http://pono.ucsd.edu/\textasciitilde{}adam/browndwarfs/spexprism/ \\
IRTF Spectral library (M.R. Cushing) \\ 
--- http://irtfweb.ifa.hawaii.edu/\textasciitilde{}spex/IRTF\_Spectral\_Library/ \\
L and T dwarf data archive (S.K. Leggett) \\ 
--- http://staff.gemini.edu/\textasciitilde{}sleggett/LTdata.html \\
NIRSPEC Brown Dwarf Spectroscopic Survey (I.S. McLean) \\
--- http://www.astro.ucla.edu/\textasciitilde{}mclean/BDSSarchive \\
Keck LRIS specra of late-M, L and T dwarfs (I.N. Reid) \\
--- http://www.stsci.edu/\textasciitilde{}inr/ultracool.html \\

\end{table*}

\subsubsection{L Dwarfs}

The L dwarf class is disinguished by the weakening and disappearance of the TiO and VO bands that are distinctive of M
dwarfs. TiO has disappeared by L6 and VO by L4. A classification scheme for L dwarfs based on the optical spectral region
(630 -- 1000 nm) is described by \citet{kirkpatrick99}. It lists spectral standards for classes L0 to L8 (see table
\ref{tab_stand}) and classification is based on the weakening TiO and VO bands, changes in CrH and FeH bands (CrH is
strongest at L5) and the alkali metals, with Cs I and Rb I lines increasing in strength to later types. 

Spectral classification of L dwarfs in the near-IR is discussed by \citet{reid01}, \citet{geballe02} and
\citet{nakajima04}. \citet{kirkpatrick10} defined a set of spectral standards for the near-IR for spectral types M0 --
L9. The near-IR region shows broad absorption bands of H$_2$O increasing in strength towards later spectral types.

While methane in the 1 -- 2.5 $\mu$m region is not seen until spectral type T, the stronger methane $\nu_3$ band in the
3.3$\mu$m region is observable in late L dwarfs \citep{noll00, schweitzer02, stephens09}. 

The physical basis for the M-L transition is thought to be the formation of condensates. At temperatures just below 2000
K the condensation of Ti bearing species such as CaTiO$_3$ (perskovite) and Ti$_2$O$_3$ removes TiO from the gas phase,
and at slightly lower temperatures VO condenses as solid VO \citep{burrows99,lodders02}. Species such as enstatite
(MgSiO$_3$), forsterite (Mg$_2$SiO$_4$), spinel (MgAl$_2$O$_4$) and solid iron also condense and these produce the dust
clouds that are necessary to explain the spectra and colours of L dwarfs \citep{allard01,marley02,tsuji02}

\subsubsection{T Dwarfs}

The T dwarf class is characterized by the appearance of methane (CH$_4$) absorption features in the near-IR region
(1--2.5 $\mu$m) . Methane first becomes apparent in early T dwarfs due to features at 1.67 and 2.2 $\mu$m which represent
the Q-branches of the strongest methane bands 2$\nu_3$ at 1.67 $\mu$m and $\nu_2 + \nu_3$ at 2.2 $\mu$m. This is
accompanied by weakening of the CO absorption at 2.3 $\mu$m.

At later types broad methane absorptions develop due to the complex methane band systems, the octad \citep[8 ground-state
bands in the 2.1 -- 2.4 $\mu$m region;][]{hilico01} and the tetradecad \citep[14 ground-state bands in the 1.6 -- 2.0
$\mu$m region;][]{nikitin13a}. These ground-state bands are associated with large numbers of hot bands. Methane
absorption is also present at around 1.4 $\mu$m (the icosad -- 20 ground-state bands) and 1.15 $\mu$m (the triacontad --
30 ground-state bands). \citet{bailey11} provides a more detailed description of the methane spectrum. 

In late T dwarfs the broad CH$_4$ and H$_2$O absorptions deepen and combine to leave a spectrum defined by approximately
triangular peaks at 1.08 $\mu$m, 1.27 $\mu$m and 1.58 $\mu$ (the ``windows'' between the deep absorptions), as well as a
weaker peak at about 2.1 $\mu$m. T dwarf spectra are also shaped by the collision-induced absorption due to H$_2$ --
H$_2$ pairs \citep{borysow02,abel11} which depresses the 2 $\mu$m peak, and by the far wings of very strong Na I and K I
lines in the optical   \citep{burrows03,allard03} which absorb at wavelengths up to $\sim$ 1 $\mu$m. 

Classification schemes for T dwarfs based on near-IR spectra, were developed by \citet{burgasser02} and \citet{geballe02}
and the two schemes were unified in \citet{burgasser06a}. That work gives a set of spectral standards for T0 -- T8 (see
table \ref{tab_stand}). The main features used for classification are the increasing depths of the H$_2$O and CH$_4$
bands towards later classes. A parallel optical classification scheme based on the 630 -- 1010 nm region is described by
\citet{burgasser03} and is based on some of the same spectral standards used in the near-IR.

The transition from L to T is associated with the switch in chemical equlibirum between CO and CH$_4$
\citep{lodders02,burrows99} that occurs at about 1400 K at 1 bar pressure, with CO being more stable above this
temperature, and CH$_4$ being favoured at lower temperatures. However, the transition is also associated with a clearing
of the dust clouds that are important in L dwarfs \citep{allard01,burgasser02}.

\subsubsection{Y Dwarfs}

The possible existence of objects even cooler than the T dwarfs was investigated in models by \citet{burrows03s}. Among
the features suggested as marking the transition to a new spectral class, were the appearance of NH$_3$ absorption, the
condensation of H$_2$O clouds, and the development of redder near-IR colours reversing the trend in T dwarfs. A number of
of very cool dwarfs were found in the CFBDS and UKIDSS surveys \citep{warren07,delorme08b,burningham08}. \citet{lucas10}
reported the discovery of an even cooler object UGPS 0722$-$05 which they suggested should be classified as T10, and
could in the future be regarded as the first example of a new spectral type.

In 2011, \citet{cushing11} reported the ``Discovery of Y-dwarfs''. Several objects identified using the WISE satellite
were found to be of later spectral types than UGPS 0722$-$05. They reclassified UGPS 0722$-$05 as the T9 spectral
standard, and classified six new objects as Y dwarfs with WISE 1738+27 as the Y0 standard. \citet{kirkpatrick12} report
several more Y dwarfs and added a spectral standard for the Y1 class (see table \ref{tab_stand}). Other reported Y dwarfs
are WISE J1639$-$68 \citep{tinney12} and the white dwarf companion WD 0806$-$661 B \citep{luhman11,luhman12}. The high
proper motion object WISE J085510.83$-$071442.5 \citep{luhman14} has absolute magnitude and colours suggesting it is the
coolest known Y dwarf with an effective temperature of 225 -- 260 K.

All Y dwarfs are very faint objects (J mag of 19 or fainter) and so the quality of available spectra are limited. They
resemble the late T dwarfs, but the ``window'' features (particularly that at 1.27 $\mu$m) become increasingly narrow
with later spectral types. The NH$_3$ absorptions expected at $\sim$1.53 and $\sim$1.03 $\mu$m are not seen at the levels
predicted by equilibrium chemistry \citep{leggett13}.

\begin{figure}
\begin{center}
\includegraphics[scale=0.44, angle=0]{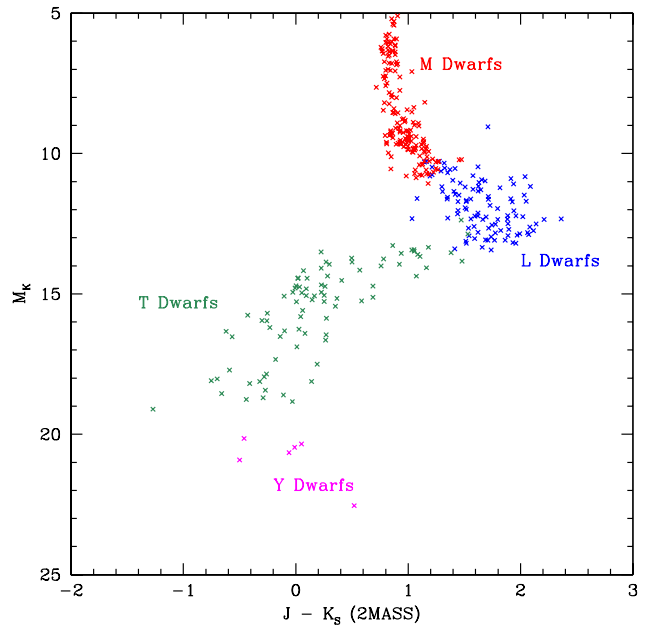}
\caption{Colour magnitude diagram (J$-$K against M$_K$) for late type dwarfs. Most of the data is taken from \citet{dupuy12}. Data on late T and Y dwarfs is from \citet{dupuy13} and has been roughly converted to the 2MASS system according to \citet{stephens04}. Additional data on earlier type M dwarfs has been added from the compilation of Reid (http://www-int.stsci.edu/\textasciitilde{}lnr/cmd.html) based on photometry from \citet{leggett92} and converted to the 2MASS system using relations in \citet{carpenter01}.} \label{fig_phot}
\end{center}
\end{figure}

\subsection{Photometry}
\label{sec_bd_phot}

Photometric data on ultracool dwarfs in the near-IR is available for a large number of objects. The database of L, T and
Y dwarfs at DwarfArchives.org, for example, lists 1281 objects most of which have JHK magnitudes. When interpreting
photometric data at JHK it is important to note that there are several different JHK systems in use. In particular the
2MASS \citep{carpenter01} and MKO \citep{simons02} systems are both widely used in brown dwarf research. The 2MASS system
uses a significantly shorter wavelength and narrower K$_s$ band compared to the K band of the MKO system. Transformations
between the systems derived from data on stars \citep{carpenter01} are unlikely to be valid for the unusual energy
distributions seen particularly in the T dwarfs. \citet{stephens04} provide a set of transformations between photometric
systems specifically for L and T dwarfs that can be used if the spectral type is known.

Much of the energy in ultracool dwarfs is in the mid infrared, and photometry for these wavelengths has become
increasingly available from Spitzer/IRAC \citep{patten06,leggett07,leggett10} and the WISE all sky catalog
\citep{wright10}.

These objects are relatively nearby and so parallax measurements of good quality are generally feasible allowing absolute
magnitudes to be derived. Conventional CCD parallax methods can be used for the earlier type objects
\citep[e.g.][]{dahn02,vrba04,andrei11}. Infrared parallaxes can be measured for the latest type objects
\citep{tinney03,dupuy12,marsh13}. The recent compilation by \citet{dupuy12} includes absolute magnitudes in the near and
mid infrared for 314 objects with known parallaxes. Mean absolute magnitudes from this compilation in the MKO JHK systems
are given in table \ref{tab_stand} supplemented by those of \citet{dupuy13} for the latest spectral types.

Figure \ref{fig_phot} shows the J$-$K against M$_K$ colour magnitude diagram for M to Y dwarfs. A distinctive feature of
the diagram is the behaviour at the L/T transition. Generally the J$-$K colour becomes slowly redder with later spectral
types through M and L, but then rapidly turn bluer through the early T spectral types. The limited photometry available
for Y dwarfs suggests a turn back to redder colours. 

In the J band a significant brightening with later spectral type can be seen \citep{dahn02,tinney03}. In the mean data of
table \ref{tab_stand} it can be seen that types L9 to T2 are all brighter at J than L6 and L7. \citet{tsuji03} suggested
that this may be an artifact of observing objects with different masses and ages, and not a feature seen in a single
cooling track. Studies of binary brown dwarfs whose components straddle the L/T boundary, however, show   ``flux
reversals'' where the cooler component is brighter in the 1 -- 1.3 $\mu$m region \citep{burgasser06b,liu06,looper08}
showing that the effect is a real intrinsic features of the L/T transition.

\subsection{Effective Temperatures}

The effective temperature of ultracool dwarfs can be determined by two main methods. The first way is to use photometry
and parallax measurements to determine the bolometric luminosity. A temperature can then be derived if the radius is
known. We don't have direct radius measurements for most of these objects, but as shown in figure \ref{fig_rad}, models
predict that the radius of brown dwarfs varies little with mass and age, so model based radius constraints can be used to
determine effective temperature.

The other way to determine effective temperatures is to fit observed spectra to those predicted by model atmospheres.
This is likely to be most reliable if the observations cover a large wavelength range that includes a substantial
fraction of the luminosity, and for brown dwarfs this means including the mid-IR as well as the near-IR
\citep[e.g.][]{stephens09}.

\begin{figure}
\begin{center}
\includegraphics[scale=0.44, angle=0]{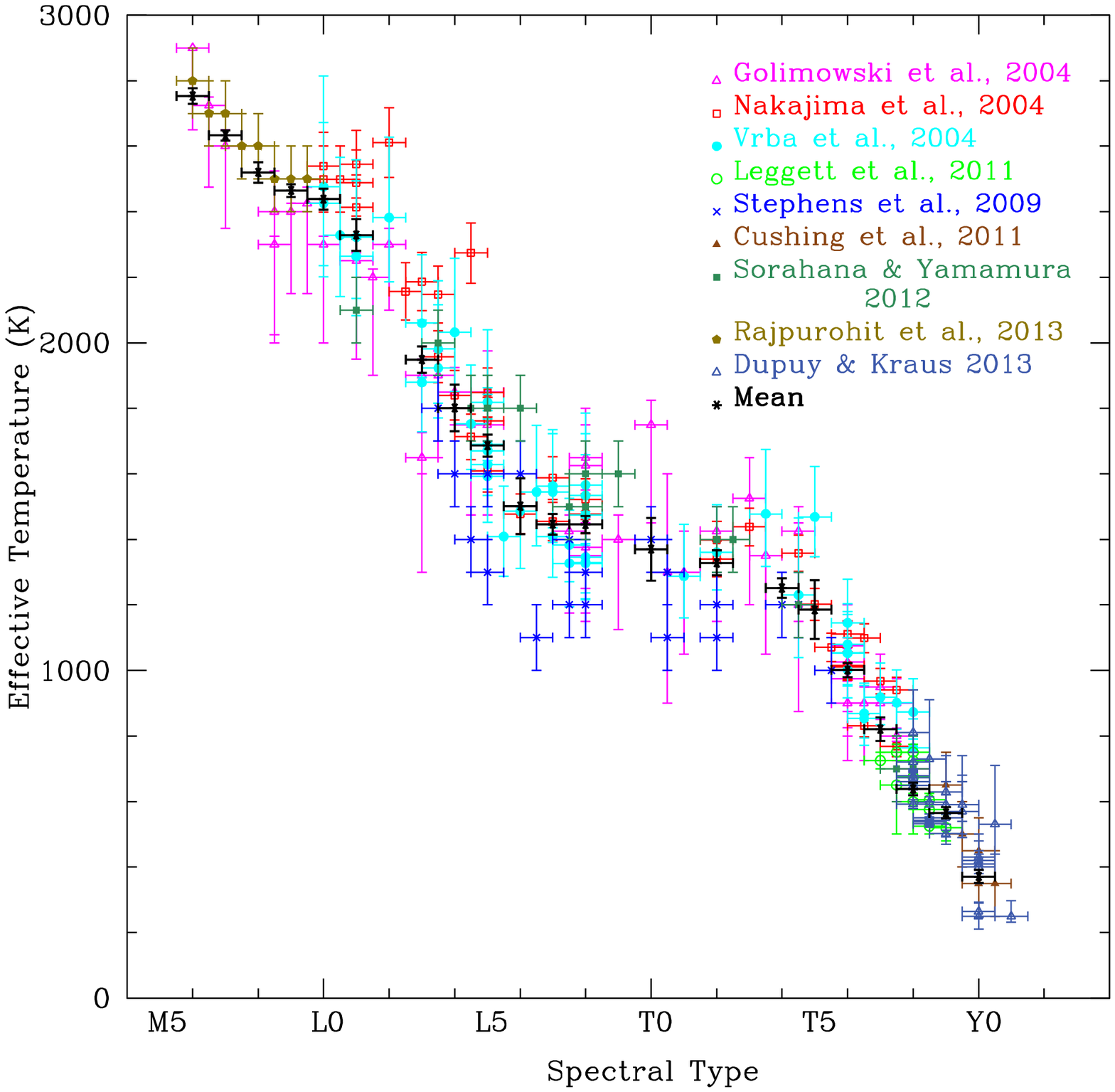}
\caption{Effective temperatures plotted against spectral type. The effective temperatures are determined from bolometric luminosities \citep{vrba04,golimowski04,nakajima04,dupuy13} or from fitting models to observed spectra \citep{stephens09, leggett11, cushing11, sorahana12, rajpurohit13}. Optical spectral types are used up to L8, and infrared spectral types for L9 and later. Late T and Y dwarf spectral types are from \citet{kirkpatrick12}. Spectral types are shown with error bars of $\pm$ 0.5 subtypes. Mean values are given for spectral types that had more than 3 measurements. Where no error estimate was given in the original publication an error bar of $\pm$100 K has been shown.} \label{fig_teff}
\end{center}
\end{figure}

Figure \ref{fig_teff} is a compilation of effective temperature measurements from the literature using both of these
methods. It shows reasonable agreement betwee the various determinations. A feature of this diagram is that, while the
general trend is decreasing temperature with later spectral type, the temperature actually changes little over the L/T
transition from about L6 to T4. This suggests that the spectral changes seen over this range are due to the clearing of
dust rather than to the direct effect of changing temperatures.

The mean effective temperatures for each spectral type from the data of figure \ref{fig_teff} have been included in table
\ref{tab_stand}.

\begin{figure} \begin{center} \includegraphics[scale=0.44, angle=0]{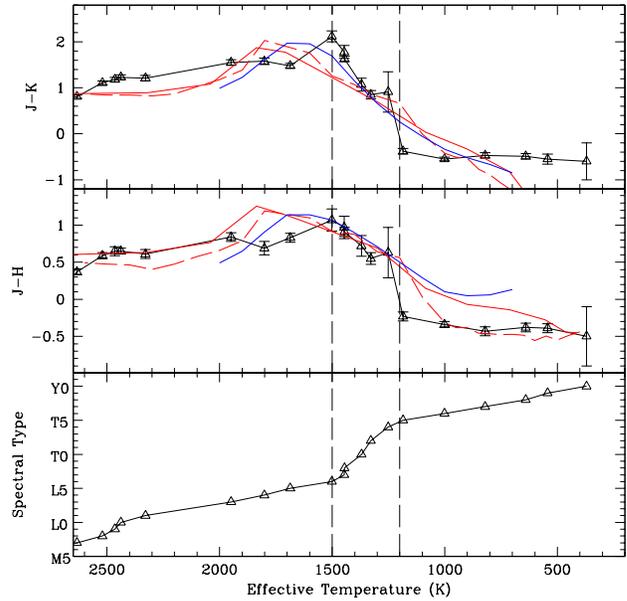} 
\caption{The L/T Transition. Over the small
range of effective temperature from 1200 -- 1500 K the observed colours and spectral types of ultracool dwarfs (black)
vary through a large range. Models (red and blue lines, see section \ref{sec_mod_clouds}) show much more gradual
changes. The data for this plot are that of table \ref{tab_stand}. Model results are from the BT-Settl models
\citep{allard07,allard12} for $\log{g}$ = 5 (solid red line) and for a 3 Gyr isochrone (red dashed line), and the Unified
Cloudy Model \citep{tsuji02,tsuji05} for $\log{g}$ = 5 and T$_{cr}$ = 1800 K (blue line).} 
\label{fig_lt_trans}
\end{center} \end{figure}

The L/T transition shows up particularly clearly when the near-IR colours and spectral types are plotted against
effective temperature using the mean values given in table \ref{tab_stand}. The J$-$K and J$-$H colours and the spectral
type all vary dramatically over the effective temperature range from 1200 -- 1500 K, and show much less variability at
other temperatures as shown in figure \ref{fig_lt_trans}. The changes are thought to be mostly due to the disappearance
of dust clouds as the atmospheres cool, but it is not clear why this should appear as such a sharp transition. Cloud models
(to be described in section \ref{sec_mod_clouds} --- the red and blue lines) show much more gradual changes than those
observed.

\subsection{Variability}
\label{sec_bd_var}

Variability has been reported in a number of L dwarfs \citep{clarke02,koen06,lane07,heinze13}. The amplitudes are
typically $\sim$1\% and the variations are quasi-periodic with periods of a few hours. The variability is generally
attributed to rotational modulation either due to patchy clouds, or magnetic spots.

Two early T dwarfs have been observed to show larger amplitude variability \citep{artigau09,radigan12}. In the case of
the T2.5 dwarf 2MASS 2139+02 an amplitude of up to 26\% was observed with a period of 7.72 hours. The large amplitude in
these early T objects is suggested to be indicative of patchy cloud regions arising during the clearing of clouds
associated with the L/T transition as suggested by \citet{marley10}. A further example reported recently \citep{gillon13}
is variability in the cooler component of the 2 pc binary brown dwarf WISE J1049$-$53 (Luhman 16).

\citet{buenzli12} have reported Spitzer and HST observations of variability with a 1.4 hr period in the T6.5 dwarf 2MASS
2228$-$43, confirming a ground-based detection of this period by \citet{clarke08}. They find phase shifts between
variations at different wavelengths which can provide a probe of the vertical atmospheric structure.

Recently \citet{crossfield14} have used time resolved near-infrared spectroscopy around the rotation period to derive a
global 2D map of the brighteness distribution of Luhman 16B using Doppler imaging techniques. The map reveals structure
that may be due to patchy clouds.

\section{EXOPLANETS}
\label{sec_giant}

\subsection{History and Properties}

Since the discovery of the first planets orbiting normal stars \citep{mayor95,marcy97} the rate of discovery has steadily
increased to more than 1800 confirmed planets according to The Extrasolar Planets Encyclopedia \citep[exoplanet.eu
---][]{schneider11} as at July 2014. In addition more than 3000 planet candidates have now been found by the Kepler
mission \citep{batalha13}. The latter are not yet confirmed planets, but it is estimated that the false positive rate for
Kepler planet candidates is likely to be $\sim$10\% \citep{morton11,fressin13}. 

While there a large number of planets, observations of their atmospheres are much more difficult than for the brown
dwarfs just considered. The vast majority of planet detections and observations are by indirect
methods, such as radial velocity measurements of the host star, and transit measurements. These provide information on
the orbit, mass and radius (for transitting planets). However, apart from a small number of directly imaged planets, we
don't yet have the capability to resolve planets from their stars in order to measure their spectra. At present most of
our data on the spectra of exoplanets comes from analysis of unresolved planets that require extracting signals that are
a small fraction of that from the host star.

A recent review of exoplanet detection methods is given by \citet{wright13}. All methods currently used are subject to
biases. The radial velocity (RV) technique that has been used for the majority of exoplanet discoveries favours the detection
of massive planets and short period orbits. The majority of RV detected planets are therefore giant planets, but at 
short periods this
method can detect planets down to a few Earth masses. When corrections are made for incompleteness the statistics show
that planet frequency increases for decreasing mass \citep{howard10,wittenmyer11}. This is consistent with the increasing
planet frequency at small sizes shown by analysis of the Kepler planet candidates \citep{howard12}. Ground-based transit
searches \citep[e.g. ][]{bakos04,pollacco06} are strongly biased toward finding large short period planets (i.e. hot
Jupiters).

Our current ability to characterise exoplanet atmospheres is largely limited limited to giant planets and to planets with
high temperatures (T $>$ $\sim$1000 K). In most cases these are hot Jupiters, i.e. massive planets that are hot because
they are close to their star, or are directly imaged massive planets that are hot because they are young planets still
cooling. There are a few cases of lower mass planets, for example Neptune/Uranus mass planets such as GJ 3470b and GJ
436b and two examples of transiting super-Earths, GJ 1214b \citep{charbonneau09} and HD 97658b for which characterization
observations have been made. 

\subsection{Solar System Giant Planets}

\begin{figure} \begin{center} \includegraphics[scale=0.43, angle=0]{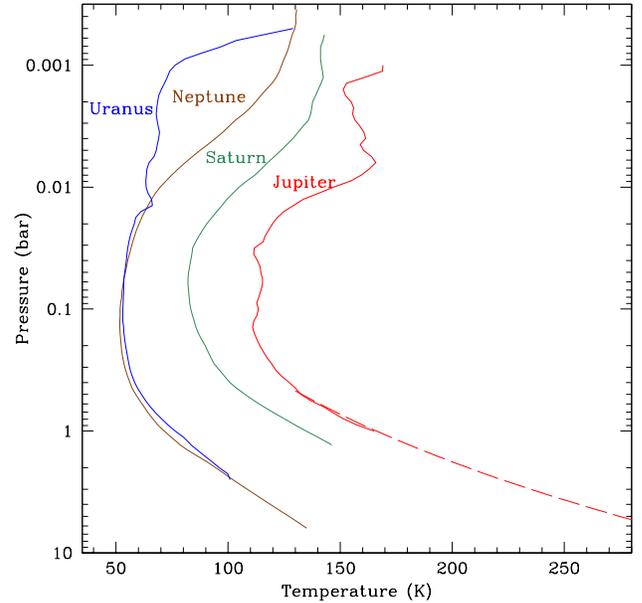} 
\caption{Temperature profiles of the Solar
system giant planet atmospheres from Voyager radio occultation measurements \citep{lindal92}, and from the Galileo probe
for Jupiter \citep[dashed line -- ][]{seiff98}.} 
\label{fig_ss_prof} 
\end{center} 
\end{figure}

\begin{figure}
\begin{center}
\includegraphics[scale=0.46, angle=0]{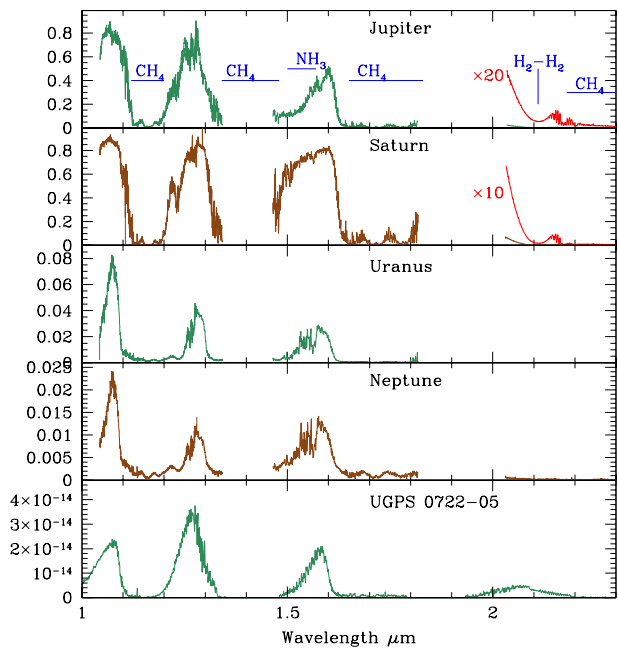}
\caption{Near-IR reflected light spectra of the Solar system giant planets Jupiter, Saturn, Uranus and Neptune 
(plotted as radiance factor I/F). The data are from IRIS2 on the Anglo-Australian Telescope as described by 
\citet{kedziora11}. The red curves show the weak K-band spectra of Jupiter and Saturn scaled up by factors of 
10 and 20. The spectrum of the T9 dwarf UGPS 0722-05 is shown for comparison using data from \citet{bochanski11}.} 
\label{fig_ss_giants}
\end{center}
\end{figure}

We do, however, know of several giant planets that have been studied in considerable detail, the giant planets in our own
Solar system. It is useful to briefly review their properties. All the giant planets have atmospheres composed of
hydrogen and helium and are enriched in heavy elements with respect to the solar composition. In the case of Jupiter
measurements with the Galileo probe show C, N, S, Ar, Kr, and Xe enriched by factors of 2 to 4 relative to solar
abundances \citep{owen99, wong04}. Carbon is enriched relative to its solar value by 7 times in Saturn \citep{flasar05}
and by 30 -- 40 times in Uranus and Neptune \citep{lodders94}.

All the Solar system giant planet atmospheres have directly measured temperature structures from radio occultation
measurements \citep{lindal92}, and from the Galileo probe \citep{seiff98} in the case of Jupiter (see figure
\ref{fig_ss_prof}). All the planets have clouds with the main cloud deck at about 0.75 bar in Jupiter
\citep{banfield98,kedziora11}, 2.5 bar in Saturn \citep{fletcher11} and $\sim$2 bar in Uranus and Neptune
\citep{irwin10}.

Near-IR spectra of the giant planets are shown in figure \ref{fig_ss_giants}. All of these are dominated by absorption
band systems due to methane (CH$_4$) centered around 1.15, 1.4, 1.7 and 2.3 $\mu$m, and are bright in the window regions
between these absorptions. In this respect the spectra resemble those of late T dwarfs, and the T9 dwarf UGPS 0722-05 is
shown in figure \ref{fig_ss_giants} for comparison. Jupiter also show absorption due to NH$_3$ at around 1.55 $\mu$m. All
the planets also show collision induced absorption due to H$_2$ -- H$_2$ pairs, which at these low temperatures shows up
as a distinctive broad feature at around 2.12 $\mu$m. This depresses the brightness in the methane window that would
otherwise be present at around 2 $\mu$m, and makes all the planets quite faint in the K-band compared with the J and H
bands.

Other species present in the atmospheres at trace levels and detected in longer wavelength spectra include PH$_3$ and
AsH$_3$ in Jupiter and Saturn \citep{fletcher11}, and hydrocarbons such as C$_2$H$_2$ and C$_2$H$_6$ in the stratospheres
\citep{hesman09,greathouse11}.

\subsection{Observing Exoplanet Atmospheres}
\label{sec_exo_obs}

\subsubsection{Direct Spectroscopy of Resolved Planets}
\label{sec_direct}

\begin{figure}
\begin{center}
\includegraphics[scale=0.44, angle=0]{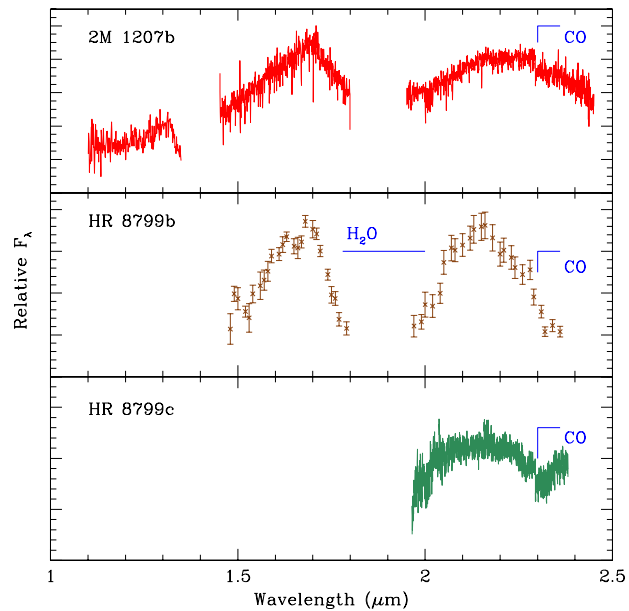}
\caption{Spectra of the direct imaged planets (or planetary mass objects) 2M 1207b \citep{patience10}, HR 8799b 
\citep{barman11a} and HR 8799c \citep{konopacky13}. The CO bandhead at 2.3 $\mu$m is apparent in all three objects 
as well as H$_2$O absorption at $\sim$1.9 and $\sim$1.4 $\mu$m.} \label{fig_direct}
\end{center}
\end{figure}

\begin{figure}
\begin{center}
\includegraphics[scale=0.44, angle=0]{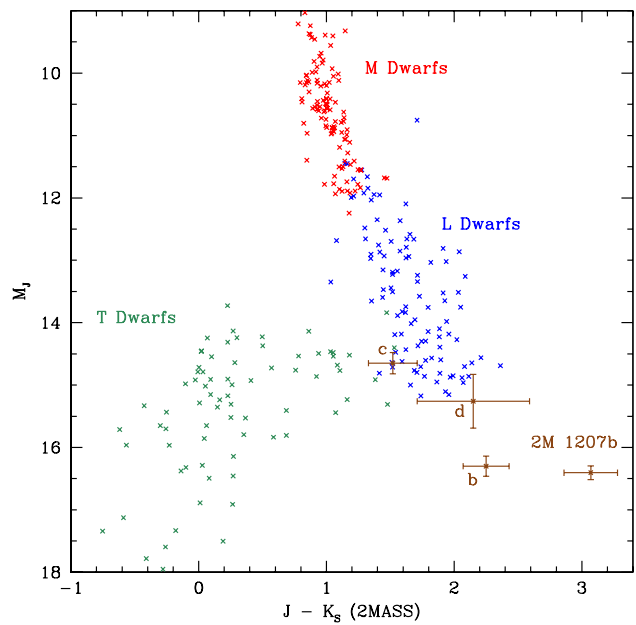}
\caption{Colour magnitude diagram for 2M 1207b and the b, c and d planets of HR 8799 
\citep[photometry from ][]{chauvin05,marois08,mohanty07} compared with field M, L and T dwarfs (from the same data 
sources as figure \ref{fig_phot}).} \label{fig_phot_direct}
\end{center}
\end{figure}

A number of ``planets'' have been discovered through direct imaging of young objects using  ground-based adaptive optics
or the Hubble Space Telescope. These include the companion of the brown dwarf 2MASSW J1207334$-$393254 \citep[usually
referred to as 2M 1207b --- ][]{chauvin05}, and the four planets of HR 8799 \citep{marois08,marois10}. 

The classification of some of these objects as planets is controversial. Although 2M 1207b was announced as the first
directly imaged extrasolar planet by its discoverers, it can be argued that it is not a planet because it orbits a brown
dwarf, not a star, or because it is unlikely that it formed through the normally understood planet formation process from
a disk around its primary object \citep{soter06}. 2M 1207b is usually referred to as a ``planetary mass object'' in
recent literature.

The classification of such objects as planets also depends on the masses determined by application of evolutionary
models, and this critically depends on the age. \citet{marois10} use age ranges from 20 -- 160 Myr for HR8799 to derive
masses for the planets in the range 5 -- 13 M$_{Jup}$ placing them most likely below the deuterium burning limit.
However, an age as high as $\sim$1 Gyr is suggested by asteroseismology methods \citep{moya10} which would make the
objects brown dwarfs rather than planets. A number of recent studies based on dynamics \citep{moro10,sudol12} and a
direct radius determination for HR 8799 \citep{baines12} favour a young age and planetary masses for the companions.

Near-IR spectra have been obtained for 2M 1207b \citep{mohanty07,patience10}, the HR 8799 planets
\citep{bowler10,barman11a,oppenheimer13,konopacky13} and $\beta$ Pic b \citep{chilcote14}. Spectra of 2M 1207b and HR 8799 b and c are shown in figure
\ref{fig_direct}. The spectra show the CO bandhead at 2.3 $\mu$m, and H$_2$O absorption at 1.4 and 1.9 $\mu$m (deepest in
HR 8799b). CH$_4$ absorption is either absent or possibly weakly present in HR 8799b. The spectral features are similar
to those of mid to late L dwarfs, which would imply objects of T$_{\mbox{\scriptsize eff}}$ $\sim$1400 -- 1600 K. 

However, photometry of 2M 1207b shows it to be very red in J$-$K and underluminous compared with L dwarfs  (figure
\ref{fig_phot_direct}). This led \citet{mohanty07} to suggest that grey extinction by an edge-on disk may be the cause of
the underluminosity. Photometry of HR 8799b show that it is similarly underluminous. \citet{barman11b} have shown that it
is possible to model the spectrum of 2M 1207b with a cool (T$_{\mbox{\scriptsize eff}}$ $\sim$1000 K) model by including
clouds and a departure from chemical equilibrium due to vertical mixing that inihbits the formation of methane. Similar
models have been fitted to the spectra of HR 8799b \citep{barman11a} and c \citep{konopacky13}.

Spectroscopy of $\beta$ Pic b in the H band \citep{chilcote14} taken with the Gemini Planet Imager shows
spectral structure indicating H$_2$O absorption and atmospheric model fits give T$_{\mbox{\scriptsize eff}}$ =
1650 $\pm$ 50 K and $\log g$ = 4.0 $\pm$ 0.25. 

A detection of methane \citep{janson13} has been reported in the planetary mass companion GJ 504b \citep{kuzuhara13}.
This was achieved using Spectral Differential Imaging with the HiCAIO adaptive optics camera on the Subaru telescope. The
companion was found to be much fainter in the CH$_4$ absorption band at $\sim$1.7 $\mu$m than in other bands indicating a
deep methane absorption comparable to that in late T dwarfs.

\subsubsection{High Resolution Cross Correlation Techniques}
\label{sec_hires}

\begin{table*}
\begin{center}
\caption{High-Resolution Cross Correlation Detections}\label{table_hires}
\begin{tabular}{lllll}
\hline\hline Planet & Species & $K_P$ & $M_P$ & Reference\\
 & Detected  & (km s$^{-1}$) & ($M_{Jup}$) & \\
\hline HD 209458b$^a$ & CO 5.6$\sigma$ & 140 $\pm$ 10 & 0.64 $\pm$ 0.09 & \citet{snellen10a} \\
$\tau$ Boo b & CO 6.2$\sigma$ & 110 $\pm$ 3.2 & 5.95 $\pm$ 0.28 & \citet{brogi12} \\ 
$\tau$ Boo b & CO 3.4$\sigma$ & 115 $\pm$ 11  & 5.6 $\pm$ 0.7 & \citet{rodler12} \\
$\tau$ Boo b  & H$_2$O 6$\sigma$   &  111 $\pm$ 5  &  5.9$_{-0.20}^{+0.35}$  & \citet{lockwood14} \\
51 Peg b     & CO/H$_2$O 5.9$\sigma$ & 134.1 $\pm$ 1.8 & 0.46 $\pm$ 0.02 & \citet{brogi13} \\
HD 189733b   & CO 5.0$\sigma$ & 154$_{-3}^{+4}$ & 1.162$_{-0.039}^{+0.058}$ & \citet{dekok13} \\
HD 189733b   & H$_2$O 4.8$\sigma$ & 154$_{-10}^{+14}$ &   &  \citet{birkby13} \\
HD 189733b   & CO 3.4$\sigma$ & 154   &   &  \citet{rodler13} \\
HD 179949b   & CO/H$_2$O 6.3$\sigma$ &  142.8 $\pm$ 3.4 &  0.98 $\pm$ 0.04 & \citet{brogi14}\\
$\beta$ Pic b &  CO 6.4$\sigma$ &     &    &  \citet{snellen14} \\
\hline\hline
\end{tabular}
\medskip\\
$^a$ Transmission spectrum during transit. All others are dayside emission detections.\\
\end{center}
\end{table*}

\begin{figure}
\begin{center}
\includegraphics[scale=0.84, angle=0]{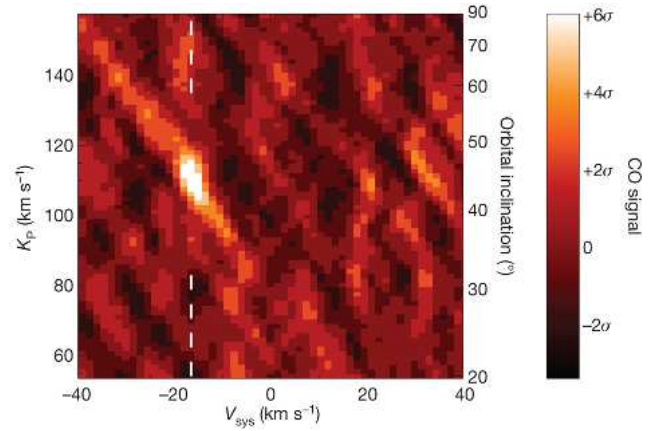}
\caption{Carbon monoxide cross correlation signal for $\tau$ Boo b \citep{brogi12} as a function of systemic velocity ($V_{sys}$) and radial velocity amplitude of the planet ($K_P$). A 6.2$\sigma$ signal is seen at $K_P = 110\pm3.2 $km s$^{-1}$ corresponding to an inclination $i = 44.5^\circ \pm$1.5 and a planet mass $M_P = 5.95\pm0.28 M_{Jup}$ ---
Reprinted by permission from Macmillan Publishers Ltd: {\it Nature}, 486, 502-504, \copyright (2012).} 
\label{fig_tauboo}
\end{center}
\end{figure}

Spectral features due to an unresolved extrasolar planet can be detected using high-resolution spectroscopy, and cross
correlation techniques to pick out the faint signal due to the planet from the much brighter contribution of the host
star. The technique was first used to attempt to detect the reflected light signal in high-resolution optical spectra of
hot Jupiters. A possible detection of a planetary signal in $\tau$ Boo was reported \citep{collier99} but was not
confirmed \citep{charbonneau99,leigh03a} and is inconsistent with the subsequent infrared detections by \citet{brogi12}
and \citet*{rodler12}. These studies set upper limits on the geometric albedo of $\tau$ Boo b of 0.3 at 480 nm
\citep{charbonneau99} and 0.39 over 400 -- 650 nm \citep{leigh03a}. Other reflected light studies for a number of the
brighter hot Jupiter systems \citep{collier02, leigh03b, rodler08, rodler10, langford11} result in similar upper limits
on geometric albedo.

Much more successful have been similar studies in the near-IR where it is possible to search for specific molecular
absorption features either in the transmission spectrum during transit \citep{snellen10a} or in the thermal emission from
the planet (which does not require a transiting planet). In these studies the telluric and stellar absorption features
are removed as best as possible and the remaining signal is cross correlated with a template spectrum. The large radial
velocity amplitude of the planet causes the absorption features to shift with orbital phase, so that a cross correlation
peak can be searched for as a function of radial velocity amplitude (K$_P$) and systemic velocity ($V_{sys}$) as shown in
figure \ref{fig_tauboo} \citep{brogi12}.

The method determines K$_P$ and thus provides a direct measurement of the planet's mass and the orbital inclination,
removing the sin $i$ uncertainty for non-transiting planets. If the planet is transiting the results can be checked
against those determined from transit analysis. The systemic velocity is also determined and should agree with that
measured for the host star. Table \ref{table_hires} list the detections reported. For two objects ($\tau$ Boo b and HD
189733b) there are independent results from two studies that are in good agreement.

Most of the objects observed in this way are hot Jupiters, but essentially the same method has also been applied to
the directly imaged exoplanet $\beta$ Pic b \citep{snellen14}. In this case it was possible to detect rotational broadening of
about 25 kms$^{-1}$ in
the CO cross correlation signal indicating a rapid rotation for the planet. 

All detections so far are either for carbon monoxide or water.  In HD189733b \citep{dekok13} CO$_2$, CH$_4$ and H$_2$O were  searched for in the 2 $\mu$m
region but not detected. However, H$_2$O was detected in HD 189733b using longer wavelength (3.2 $\mu$m) observations
\citep{birkby13}. While CO is expected to be strong feature in these planets, part of the reason it is most easily
detected may be that as a diatomic molecule it has a simpler spectrum and better quality line lists. Difficulty in
detecting other species may, in part, be due to errors in the template spectra due to problems with the line lists, such
as errors in line positions \citep[see discussion in][]{barnes10} and incompleteness. Methane line lists used
for atmospheric modelling are known to be missing many hot bands that are needed at the high temperatures of these
objects.

\subsubsection{Secondary Eclipse Photometry and Spectroscopy}
\label{sec_occn}

The secondary eclipse (or occultation\footnote{Referring to this event as the secondary eclipse is consistent with
standard terminology for eclipsing binary systems. The term occultation for this event is suggested by terminology used
in the Solar system for e.g. the phenomena of Jupiter's satellites, where eclipses, occultations and transits all occur,
and eclipse is reserved for the case where a satellite passes into the shadow of the planet. Both terms are used in the
exoplanet literature with secondary eclipse being more common.}) occurs when a planet passes behind the star. If the
planet is sufficiently bright a measurable dip in the light curve is seen, and the fractional depth of the dip is a
direct measurement of the flux from the planet as a fraction of that from the star. In most cases such measurements
detect thermal emission from the dayside of the planet, and so contrasts are greatest at infrared wavelengths.

The first detection of a secondary eclipse was made at 24 $\mu$m for HD 209458b using the Spitzer Space Telescope
\citep{deming05}. Since then a substantial number of mostly hot Jupiter type systems have had their secondary eclipse
depth measured in the Spitzer/IRAC bands (3.6 $\mu$m, 4.5 $\mu$m, 5.8 $\mu$m and 8.0 $\mu$m). There are also a number of
measurements at shorter wavelengths from ground-based telescopes. The broad band eclipse depth results are summarized in
table \ref{table_occ}. This lists secondary eclipse depths measured in the four Spitzer/IRAC bands and in the K$_s$ band
(2.15 $\mu$m). Where there are multiple measurements in a band the one with the smaller quoted error is listed, but
references to all measurements are given. The ``Other'' column lists other bands in which eclipse depths have been
measured and the references to these are also given.

\begin{table*}
\begin{center}
\caption{Secondary eclipse depth broadband measurements (\%) }\label{table_occ}
\small
\begin{tabular}{lp{1.5cm}p{1.5cm}p{1.5cm}p{1.5cm}p{1.5cm}ll}
\hline\hline Planet & K$_s$ & 3.6$\mu$m & 4.5$\mu$m & 5.8 $\mu$m & 8.0 $\mu$m & Other & References \\
\hline CoRoT-1b & 0.336$\pm$.042 & 0.415$\pm$.042 & 0.482$\pm$.042 & & & 0.71$\mu$m & 1,2,3,4 \\
CoRoT-2b & & 0.355$\pm$.020 & 0.500$\pm$.020 & & 0.510$\pm$.059 & 0.71$\mu$m & 5,6,7 \\
 GJ436b & & 0.041$\pm$.003 & $<$0.010 & 0.033$\pm$.014 & 0.054$\pm$.008 & 16, 24 $\mu$m & 8,58 \\
 HD 149026b & & 0.040$\pm$.003 & 0.034$\pm$.006 & 0.044$\pm$.010 & 0.052$\pm$.006 & 16$\mu$m & 9 \\
 HD 189733b & & 0.147$\pm$.004 & 0.179$\pm$.004 & 0.310$\pm$.034 & 0.344$\pm$.004 & 16, 24$\mu$m & 10,11,12,59 \\
 HD 209458b & &0.094$\pm$.009 & 0.139$\pm$.007 & 0.301$\pm$.043 & 0.240$\pm$.026 & 24$\mu$m & 13,14,15,62 \\
 HAT-P-1b & 0.109$\pm$.025& 0.080$\pm$.008 & 0.135$\pm$.022 & 0.203$\pm$.031 & 0.238$\pm$.040 & & 16,17 \\
 HAT-P-3b & & 0.112$_{-.030}^{+.015}$ & 0.094$_{-.009}^{+.016}$ & & & & 18 \\
 HAT-P-4b & & 0.142$_{-.016}^{+.014}$ & 0.122$_{-.014}^{+.012}$ & & & & 18 \\
 HAT-P-6b & & 0.117$\pm$.008 & 0.106$\pm$.006 & & & & 19 \\
 HAT-P-7b & & 0.098$\pm$.017 & 0.159$\pm$.022 & 0.245$\pm$.031 & 0.225$\pm$.052 & & 20 \\
 HAT-P-8b & & 0.131$_{-.010}^{+.007}$ & 0.111$_{-.007}^{+.008}$ & & & & 19 \\
 HAT-P-23b & 0.234$\pm$.046 & 0.248$\pm$.019 & 0.309$\pm$.026 & & & & 60 \\
 Kepler-5b & & 0.103$\pm$.017 & 0.107$\pm$.015 & & & & 21 \\
 Kepler-6b & & 0.069$\pm$.027 & 0.151$\pm$.019 & & & & 21 \\
 Kepler-12b & & 0.137$\pm$.020 & 0.116$\pm$.031 & & & & 22 \\
 TrES-1b & & & 0.066$\pm$.013 & & 0.225$\pm$.036 & & 23 \\
 TrES-2b & 0.062$_{-.011}^{+.013}$ & 0.127$\pm$.021 & 0.230$\pm$.024 & 0.199$\pm$.054 & 0.359$\pm$.060 & & 24,25 \\
 TrES-3b & 0.133$_{-0.016}^{+0.018}$ & 0.346$\pm$.035 & 0.372$\pm$.054 & 0.449$\pm$.097 & 0.475$\pm$.046 & & 26,27,28 \\ TrES-4b &  & 0.137$\pm$.011 & 0.148$\pm$.016 & 0.261$\pm$.059 & 0.318$\pm$.044 & & 29 \\
 WASP-3b & 0.181$\pm$.020 & 0.209$_{-0.028}^{+0.040}$ & 0.282$\pm$0.012 & & 0.328$_{-0.055}^{+0.086}$ & & 30,64 \\
 WASP-4b & 0.182$_{-.013}^{+.014}$ & 0.319$\pm$.031 & 0.343$\pm$.027 & & & & 31,32 \\
 WASP-5b & 0.269$\pm$.062 & 0.197$\pm$.028 & 0.227$\pm$.025 & & & H & 33,63 \\
 WASP-8b & & 0.113$\pm$.018 & 0.069$\pm$.007 & 0.093$\pm$.023 & & & 34 \\
 WASP-12b & 0.339$\pm$.014$^a$ & 0.419$\pm$.014$^b$ & 0.424$\pm$.021$^b$ & 0.694$\pm$.057$^b$ & 0.701$\pm$.074$^b$ & z, J, H, 2.3$\mu$m & 35,36,37,38,39,40 \\
 WASP-14b & & 0.19$\pm$0.01 & 0.224$\pm$.018 & & 0.181$\pm$0.022 & & 61 \\
 WASP-17b & & & 0.229$\pm$.013 & 0.237$\pm$.039 & & & 41 \\
 WASP-18b & & 0.304$\pm$.017 & 0.379$\pm$.008 & 0.37$\pm$.03 & 0.41$\pm$.02 & & 42,43 \\
 WASP-19b & 0.287$\pm$.020 & 0.483$\pm$.025 & 0.572$\pm$.030 & 0.65$\pm$.11 & 0.73$\pm$.12 & z, H & 44,45,46,47,57,67 \\
WASP-24b  & & 0.159$\pm$.013 & 0.202$\pm$.018 & & & & 48 \\	  
WASP-26b  & & 0.126$\pm$.013 & 0.149$\pm$.016 & & & & 49\\
WASP-33b  & 0.244$_{-.020}^{+.027}$ & 0.26$\pm$.05 & 0.41$\pm$.02 & & & z & 50,51,52 \\
WASP-43b  & 0.181$\pm$.027 & 0.346$\pm$.013 & 0.382$\pm$.015 & & & & 53,66,67 \\
WASP-46b  & 0.253$_{-.060}^{+.063}$ &	 &  &  &  &  & 65 \\
WASP-48b  & 0.109$\pm$.027 & 0.176$\pm$.013 & 0.214$\pm$.020 & & & & 60 \\
XO-1b     & & 0.086$\pm$.007 & 0.122$\pm$.009 & 0.261$\pm$.031 & 0.210$\pm$.029 & & 54 \\
XO-2b     & & 0.081$\pm$.017 & 0.098$\pm$.020 & 0.167$\pm$.036 & 0.133$\pm$.049 & & 55 \\
XO-3b     & & 0.101$\pm$.004 & 0.143$\pm$.006 & 0.134$\pm$.049 & 0.150$\pm$.036 & & 56 \\
XO-4b     & & 0.056$_{-.006}^{+.012}$ & 0.135$_{-.007}^{+.010}$ & & & & 19 \\
\hline\hline
\end{tabular}
\end{center}
\medskip
\small
$^a$Value from \citet{croll11} as corrected by \citet{crossfield12b} \\
$^b$Values from \citet{campo11} as corrected by \citet{crossfield12b} \\

{\bf References}
\small
1. \citet{deming11}, 2. \citet{gillon09}, 3. \citet{rogers09}, 4. \citet{snellen09}, 5. \citet{deming11},
6. \citet{gillon10}, 7. \citet{snellen10b}, 8. \citet{deming07}, 9. \citet{stevenson12}, 10. \citet{charbonneau08},
11. \citet{knutson12}, 12. \citet{deming06}, 13. \citet{deming05}, 14. \citet{knutson08}, 15. \citet{crossfield12a},
16. \citet{todorov10}, 17. \citet{demooij11}, 18. \citet{todorov13}, 19. \citet{todorov12}, 20 \citet{christiansen10},
21. \citet{desert11a}, 22. \citet{fortney11}, 23. \citet{charbonneau05}, 24. \citet{odonovan10}, 25. \citet{croll10a},
26. \citet{fressin10}, 27. \citet{demooij09}, 28. \citet{croll10b}, 29. \citet{knutson09a}, 30. \citet{zhao12a},
31. \citet{beerer11}, 32. \citet{caceres11}, 33. \citet{baskin13}, 34. \citet{cubillas13}, 35. \citet{campo11},
36. \citet{croll11}, 37. \citet{crossfield12b}, 38. \citet{zhao12b}, 39. \citet{cowan12b}, 40. \citet{lopez10},
41. \citet{anderson11}, 42. \citet{nyemeyer11}, 43. \citet{maxted13}, 44. \citet{anderson13}, 45. \citet{gibson10},
46. \citet{lendl13}, 47. \citet{burton12}, 48. \citet{smith12}, 49. \citet{mahtani13}, 50. \citet{deming12},
51. \citet{demooij13}, 52. \citet{smith11}, 53. \citet{blecic14}, 54. \citet{machalek08}, 55. \citet{machalek09},
56. \citet{machalek10}, 57. \citet{zhou13}, 58. \citet{stevenson10}, 59. \citet{agol10}, 60. \citet{orourke13},
61. \citet{blecic13}, 62. \citet{zellem14},   63.  \citet{chen14a},   64.  \citet{rostron14},   65. \citet{chen14b},
66. \citet{wang13}, 67. \citet{zhou14}
\end{table*}

\begin{table*}
\begin{center}
\caption{Dayside emission spectroscopy from secondary eclipses}\label{table_occ_spec}
\begin{tabular}{lllll}
\hline\hline Planet & Wavelengths & Instrument & Features Reported & Reference \\
\hline  CoRoT-2b & 1.1 -- 1.7 $\mu$m &  HST/WFC3  &  &  \citet{wilkins14}  \\
HD 189733b & 5 -- 14 $\mu$m & Spitzer/IRS & H$_2$O & \citet{grillmair08} \\
      & 1.5 -- 2.5 $\mu$m & HST/NICMOS  &  H$_2$O, CO, CO$_2$ & \citet{swain09a} \\
      & 2.0 -- 4.1 $\mu$m & IRTF/SpeX & Emission feature at 3.3 $\mu$m & \citet{swain10} \\
      & 3.27 -- 3.31 $\mu$m & Keck/NIRSPEC & No emission feature at 3.3 $\mu$m & \citet{mandell11} \\
      & 2.0 -- 4.1 $\mu$m & IRTF/SpeX & Confirm emission at 3.3 $\mu$m & \citet{waldmann12} \\
      & 1.1 -- 1.7 $\mu$m & HST/WFC3 & H$_2$O    & \citet{crouzet14} \\
HD 209458b & 7.5 -- 13.2 $\mu$m & Spitzer/IRS &  & \citet{richardson07} \\
    & 1.5 -- 2.5 $\mu$m & HST/NICMOS & H$_2$O, CH$_4$, CO$_2$ & \citet{swain09b} \\
TrES-3b &  1.1 -- 1.7 $\mu$m   &  HST/WFC3  &   &   \citet{ranjan14} \\     
WASP-4b &  1.1 -- 1.7 $\mu$m   &  HST/WFC3  &   &   \citet{ranjan14} \\   
WASP-12b  &  1 -- 2.5 $\mu$m & IRTF/SpeX & & \citet{crossfield12c} \\
          &  1.1 -- 1.7 $\mu$m & HST/WFC3 & & \citet{swain13} \\
WASP-19b  &  1.25 -- 2.35 & Magellan/MMIRS & & \citet{bean13} \\
WASP-43b  &  1.1 -- 1.7 $\mu$m  & HST/WFC3  & H$_2$O  &  \citet{kreidberg14b} \\

\hline\hline
\end{tabular}
\medskip\\

\end{center}
\end{table*}

\begin{figure}
\begin{center}
\includegraphics[scale=0.44, angle=0]{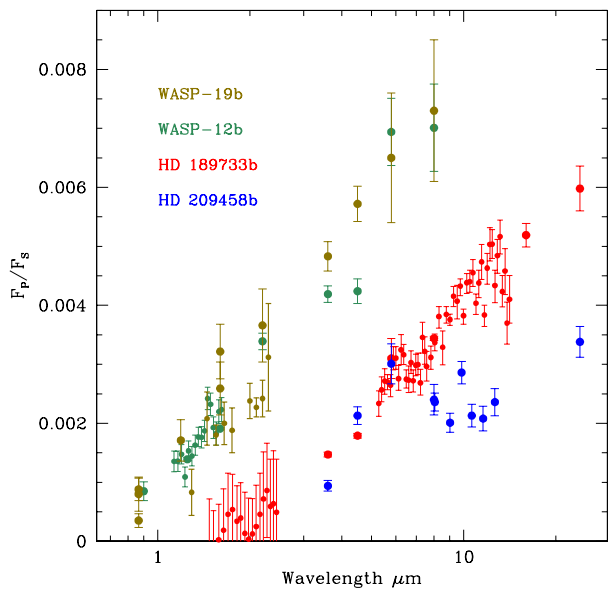}
\caption{Dayside emission from WASP-12b, WASP-19b, HD189733b and HD209458b based on data listed in tables \ref{table_occ} and \ref{table_occ_spec}. Large symbols are broad band measurements and smaller symbols are spectroscopic observations.} 
\label{fig_dayside}
\end{center}
\end{figure}

For a few of the brighter systems it is possible to go further and obtain spectra of the dayside emission using the
secondary eclipse depth. Such results are listed in table \ref{table_occ_spec}. Figure \ref{fig_dayside} shows the
combined data from broad band and spectroscopic observations for some of the best studied cases.

With secondary eclipse data of sufficient quality it is possible to map the brightness distribution across the disk of
the planet \citep{williams06}. This has been attempted for HD 189733b by \citet{majeau12} and \citet{dewit12}. The
results show a bright spot shifted east from the subsolar point in agreement with results from the full phase light curve
\citep{knutson07a}

While the infrared secondary eclipse shows the thermal emission from the planet, observations of the secondary eclipse at
visible wavelengths can show the planet through light reflected from its star. However, if the planet is very hot,
thermal emission may still be present even at visible wavelengths. Table \ref{table_geom_alb} summarises measurements so
far, mostly from observations with Kepler, in a broad band covering 400 -- 850 nm. These observations provide a measure
of the geometric albedo of the planet, and show that some of these planets are quite dark, while others have geometric
albedos up to $\sim$0.4. In the case of HD 189733b observations have been made with STIS showing the planet to be dark at
450 -- 570 nm, but with an albedo of 0.4 at 290 -- 450 nm, the blue colour being indicative of a Rayleigh scattering haze
\citep{evans13}. Low albedos in the visible are to be expected for clear atmospheres due to the broad sodium and
absorption lines, whereas higher albedos can result if clouds are present \citep{sudarsky00}.

\begin{table*}
\begin{center}
\caption{Geometric Albedo Measurements from Optical Secondary Eclipses}\label{table_geom_alb}
\begin{tabular}{llll}
\hline\hline Planet & Secondary Eclipse & Geometric & Reference \\
     & Depth (ppm) & Albedo & \\
\hline HD 189733b (290--450 nm) &  126$^{+37}_{-36}$   &  0.40$^{+0.12}_{-0.11}$ & \citet{evans13} \\
HD 189733b (450--570 nm) & 1$^{+37}_{-30}$  &  0.00$^{+0.12}_{-0.10}$ & \citet{evans13} \\
HD 209458b & 7 $\pm$ 9 &  0.038 $\pm$ 0.045  &    \citet{rowe08} \\
Kepler-5b & 21 $\pm$ 6 & 0.12 $\pm$ 0.04 & \citet{desert11a} \\
          & 19 $\pm$ 4 & 0.12 $\pm$ 0.02 & \citet{esteves13} \\
	  & 19.8 $\pm$ 3.65  &  0.16 $\pm$ 0.03  &  \citet{angerhausen14} \\
Kepler-6b & 22 $\pm$ 7 & 0.11 $\pm$ 0.04 & \citet{desert11a} \\
          & 8.9 $\pm$ 3.8 & 0.059 $\pm$ 0.025 & \citet{esteves13} \\
	  & 11.3 $\pm$ 4.2  &  0.07 $\pm$ 0.03  &  \citet{angerhausen14} \\
Kepler-7b & 47 $\pm$ 14 & 0.38 $\pm$ 0.12 & \citet{kipping11} \\
          & 42 $\pm$ 4 & 0.32 $\pm$ 0.03 & \citet{demory11} \\
	  & 48 $\pm$ 3 & 0.35 $\pm$ 0.02 & \citet{demory13} \\
	  & 46.6 $\pm$ 4.0  &  0.32 $\pm$ 0.03  & \citet{angerhausen14} \\
Kepler-8b & 26 $\pm$ 6 & 0.14 $\pm$ 0.03 & \citet{esteves13} \\
          & 16.5 $\pm$ 4.45  & 0.11 $\pm$ 0.03  &  \citet{angerhausen14} \\
Kepler-12b & 31 $\pm$ 7 & 0.14 $\pm$ 0.04 & \citet{fortney11} \\
           & 18.7 $\pm$ 4.9  &  0.09 $\pm$ 0.02  & \citet{angerhausen14} \\
Kepler-13Ab &  173.7 $\pm$ 1.8 &  0.33$_{-0.06}^{+0.04}$ & \citet{shporer14}  \\
Kepler-17b & 58 $\pm$ 10 & 0.10 $\pm$ 0.02 & \citet{desert11b} \\
           &  43.7 $\pm$ 6.4 &  0.08 $\pm$ 0.01 & \citet{angerhausen14} \\
Kepler-41b & 64$^{+10}_{-12}$ & 0.30 $\pm$ 0.08 & \citet{santerne11} \\
           & 60 $\pm$ 9 & 0.23 $\pm$ 0.05 & \citet{quintana13} \\
	   & 46.2 $\pm$ 8.7  &  0.18 $\pm$ 0.03  &  \citet{angerhausen14}  \\
Kepler-43b &  17.0 $\pm$ 5.3 &  0.06 $\pm$ 0.02  &  \citet{angerhausen14}  \\
Kepler-76b &  75.6 $\pm$ 5.6 &  0.22 $\pm$ 0.02  &  \citet{angerhausen14}  \\	
Kepler-412b & 47.4 $\pm$ 7.4 &  0.094 $\pm$ 0.015 &  \citet{deleuil14} \\
           & 40.2 $\pm$ 9.0  &  0.11 $\pm$ 0.02  &  \citet{angerhausen14} \\   
TrES-2b    & 6.5 $\pm$ 1.9 & 0.025 $\pm$ 0.007 & \citet{kipping11s} \\
           & 7.5 $\pm$ 1.7 & 0.030 $\pm$ 0.007 & \citet{esteves13} \\

\hline\hline \end{tabular} \medskip\\ All measurements are with Kepler (400 -- 850 nm) except for HD 209458b observed
with MOST (350 -- 700 nm) and HD 189733b observed with HST/STIS. \end{center} \end{table*}

\subsubsection{Transit Spectroscopy}
\label{sec_transit}

\begin{table*}
\begin{center}
\caption{Transmission Spectroscopy during Transit}\label{table_trans}
\begin{tabular}{lllll}
\hline\hline Planet & Wavelengths & Instrument & Featues Reported & Reference \\
\hline  CoRoT-1b  &  0.8 -- 2.4 $\mu$m &  IRTF/SpeX/MORIS & no TiO/VO & \citet{schlawin14} \\
     &  1.1 -- 1.7 $\mu$m  &  HST/WFC3 &   &  \citet{ranjan14} \\
GJ 436b & 1.1 -- 1.7 $\mu$m & HST/WFC3 &   & \citet{knutson14a} \\
GJ 1214b & 0.78 -- 1.00 $\mu$m & VLT/FORS1 &    & \citet{bean10} \\
             & 2.1 -- 2.4 $\mu$m & Keck/NIRSPEC &     & \citet{crossfield11} \\
             & 1.1 -- 1.7 $\mu$m & HST/WFC3 &       & \citet{berta12} \\
	     & 1.1 -- 1.7 $\mu$m & HST/WFC3 &  clouds   & \citet{kreidberg14a} \\
GJ 3470b     & 2.09 -- 2.36 $\mu$m & Keck/MOSFIRE &    & \citet{crossfield13} \\
        &  1.1 -- 1.7 $\mu$m  &  HST/WFC3  &       &  \citet{ehrenreich14}   \\	
HD 97658b & 1.1 -- 1.7 $\mu$m  &  HST/WFC3  &    &   \citet{knutson14b}  \\     
HD 189733b & 1.4 -- 2.5 $\mu$m & HST/NICMOS & H$_2$O, CH$_4$ &  \citet{swain08} \\
             & 0.55 -- 1.05 $\mu$m & HST/ACS & haze   &  \citet{pont08} \\
	     & 1.66 -- 1.87 $\mu$m & HST/NICMOS & haze, no H$_2$O &   \citet{sing09} \\
	     & 0.29 -- 0.57 $\mu$m & HST/STIS  & haze   & \citet{sing11a} \\
	     & 1.1 -- 1.7  $\mu$m &  HST/WFC3  &        & \citet{gibson12} \\
	     & 1.1 -- 1.7  $\mu$m &  HST/WFC3  &  H$_2$O      & \citet{mccullough14} \\
HD 209458b   & 0.3 -- 1.0 $\mu$m &   HST/STIS  & H$_2$O$^a$ & \citet{knutson07b} \\
	     & 1.1 -- 1.7 $\mu$m &   HST/WFC3  & H$_2$O   & \citet{deming13}  \\
HAT-P-1b     & 1.1 -- 1.7 $\mu$m &   HST/WFC3  & H$_2$O   & \citet{wakeford13} \\
HAT-P-12b    & 1.1 -- 1.7 $\mu$m &   HST/WFC3  & clouds   & \citet{line13a} \\
             & 0.52 -- 0.93 $\mu$m & Gemini/GMOS &        & \citet{gibson13a} \\
TrES-2b  &  1.1 -- 1.7 $\mu$m  &  HST/WFC3 &   &   \citet{ranjan14}  \\ 	     
TrES-4b  &  1.1 -- 1.7 $\mu$m  &  HST/WFC3 &   &   \citet{ranjan14}  \\ 	
WASP-6b      & 0.48 -- 0.86 $\mu$m & Baade/IMACS &        & \citet{jordan13} \\
WASP-12b     & 1.1 -- 1.7 $\mu$m &   HST/WFC3  &          & \citet{swain13} \\
             & 0.7 -- 1.0 $\mu$m &   Gemini/GMOS &        & \citet{stevenson14} \\
	     & 0.3 -- 1.7 $\mu$m &   HST/STIS+WFC3 & aerosols, no TiO & \citet{sing13} \\
	     & 1.1 -- 1.7 $\mu$m &   HST/WFC3      &      & \citet{mandell13} \\
WASP-17b     & 1.1 -- 1.7 $\mu$m &   HST/WFC3      &      & \citet{mandell13} \\	     
WASP-19b     & 1.25 -- 2.35 $\mu$m & Magellan/MMIRS &     & \citet{bean13} \\
             & 0.29 -- 1.69 $\mu$m & HST/STIS+WFC3 & H$_2$O no TiO & \citet{huitson13} \\
	     & 1.1 -- 1.7 $\mu$m   & HST/WFC3    &        & \citet{mandell13} \\
WASP-29b     & 0.51 -- 0.72 $\mu$m & Gemini/GMOS &         & \citet{gibson13b} \\
WASP-43b  &  0.52 -- 1.04 $\mu$m  &  GTC/OSIRIS & & \citet{murgas14}  \\
          &  1.1 -- 1.7 $\mu$m  & HST/WFC3  & H$_2$O  &  \citet{kreidberg14b} \\
XO-1b        & 1.2 -- 1.8 $\mu$m  &  HST/NICMOS & H$_2$O, CO$_2$, CH$_4$  &  \citet{tinetti10} \\
             & 1.1 -- 1.7 $\mu$m  &  HST/WFC3    &  H$_2$O   &  \citet{deming13} \\
XO-2b        & 1.1 -- 1.7 $\mu$m  &  HST/NICMOS  &      &  \citet{crouzet12} \\	     
\hline\hline
\end{tabular}
\medskip\\
$^a$ According to reanalysis by \citet{barman07}. \\
\end{center}
\end{table*}

Observations during transit (or primary eclipse, when the planet passes in front of the star) also provide information on
the atmospheres. The depth of the primary eclipse is a measure of the radius of the planet, and will be larger where
absorption is strongest.

\begin{table*}
\begin{center}
\caption{Sodium (589nm) detections from transit spectroscopy}\label{table_na}
\begin{tabular}{lllll}
\hline\hline Planet & Instrument & Line Depth & Band & Reference \\
\hline 	     
HD 189733b & HET/HRS & $6.72\pm2.07 \times 10^{-4}$ &  1.2 nm & \citet{redfield08} \\
           & HET/HRS & $5.26\pm1.69 \times 10^{-4}$ &   1.2 nm  & \citet{jensen11} \\
	   & HST/STIS & $9\pm1 \times 10^{-4}$ & 0.5 nm & \citet{huitson12} \\
HD 209458b & HST/STIS & $2.32\pm0.57 \times 10^{-4}$ &   1.2 nm & \citet{charbonneau02} \\
           & HET/HRS & $2.63\pm0.81 \times 10^{-4}$ &  1.2 nm & \citet{jensen11} \\
	   & Subaru/HDS & $5.6\pm0.7^a \times 10^{-4}$ &  0.3 nm & \citet{snellen08} \\
	   & HST/STIS & $11 \times 10^{-4}$ &  0.44 nm & \citet{sing08} \\
HAT-P-1b  &  HST/STIS & $9.8\pm3.0 \times 10^{-4}$  & 3.0 nm &  \citet{nikolov14}  \\
WASP-17b   & VLT/GIRAFFE & $55\pm13 \times 10^{-4}$ & 0.15 nm & \citet{wood11} \\
           & Magellan/MIKE & $58\pm13 \times 10^{-4}$ &  0.15 nm & \citet{zhou12} \\   
XO-2b &  GTC/OSIRIS &  $4.7\pm1.1 \times 10^{-4}$ &  5.0 nm  & \citet{sing12}  \\
\hline\hline
\end{tabular}
\medskip\\
a $7.0\pm1.1 \times 10^{-4}$ (0.15 nm band), $13.5\pm1.7 \times 10^{-4}$ (0.075 nm band)
\end{center}
\end{table*}

Spectroscopy during transits can reveal absorption features in the transmission spectrum of the planet's atmosphere.
Transit spectroscopy samples the terminator of the planet and the long tangent path length means that it is sensitive to
higher levels in the atmosphere than dayside emission spectroscopy from secondary eclipses.

\begin{figure}
\begin{center}
\includegraphics[scale=0.61, angle=0]{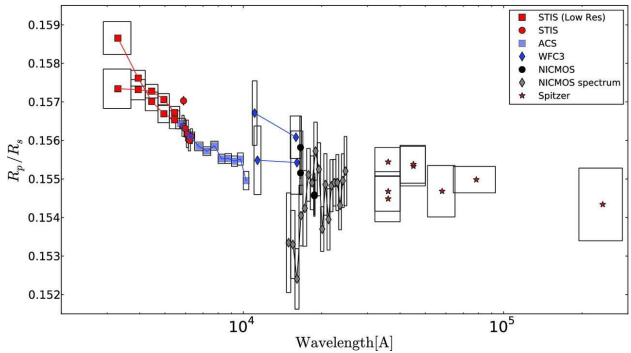}
\caption{Transmission spectrum of HD 189733b from HST and Spitzer transit observations, showing increase to the blue 
interpreted as due to a Rayleigh scattering haze. ---  Figure 9 from ``The prevalence of dust on the exoplanet 
HD189733b from Hubble and Spitzer observations'', Pont, F. et al., 2013, MNRAS, 432, 2917.} 
\label{fig_hd_trans}
\end{center}
\end{figure}

The first detection of an exoplanet atmosphere was in observations of the transits of HD 209458 \citep{charbonneau02}
that showed absorption in the sodium doublet at 589.3 nm. Transit spectroscopy (other than studies of Na line absorption)
studies are listed in table \ref{table_trans}. In addition to these spectroscopy observations there are numerous transit
measurements in broad band filters, including measurements in the Spitzer/IRAC bands that extend coverage to longer
wavelengths. HD 189733b is a particularly well studied system, and the various space observations have been combined by
\citet{pont13} to give the transmission spectrum shown in figure \ref{fig_hd_trans}. It shows increasing absorption to
the blue indicating the presence of a Rayleigh scattering haze. Wasp 12b \citep{sing13} also shows a similar increase to
the blue attributed to Rayleigh scattering from aerosols.

\begin{figure}
\begin{center}
\includegraphics[scale=0.44, angle=0]{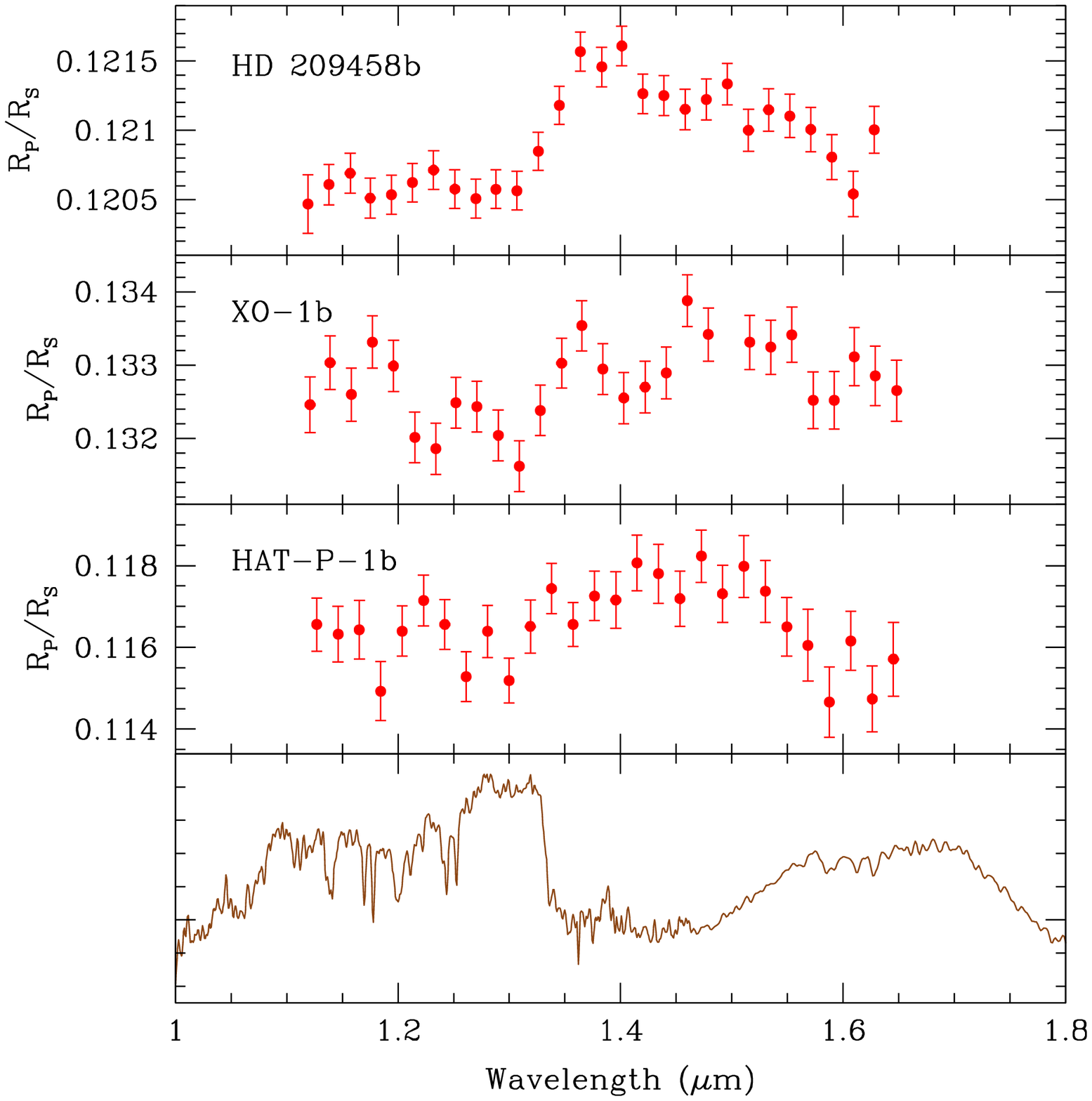}
\caption{HST/WFC3 observations of the transmission spectra of HD209458b and XO-1b \citep{deming13} and HAT-P-1b 
\citep{wakeford13} showing absorption at $\sim$1.4 $\mu$m attributed to H$_2$O. The spectrum of the brown dwarf 
Kelu-1 showing the same absorption feature is shown for comparison in the bottom panel. The exoplanet spectra are 
inverted compared with the brown dwarf spectra since absorption features increase the radius of the planet as seen 
in transit observations} 
\label{fig_wfc3_trans}
\end{center}
\end{figure}

Transmission spectra in the near-IR for three systems are shown in figure \ref{fig_wfc3_trans}, showing water vapour
absorption at $\sim$1.4 $\mu$m. As is conventional, these observations are plotted as R$_P$/R$_S$ (i.e. the radius of the
planet divided by the radius of the star as determined from the transit). This gives an inverted spectrum compared with
conventional spectroscopy, since absorption features increase the apparent radius of the planet.

Measurements of the Sodium D-line absorption are listed in table \ref{table_na}. Results are listed here where the
absorption is detected at greater than the 3-sigma level. There are also a number of unsuccesful attempts at detections.
Potassium absorption has been reported in XO-2b \citep{sing11b}.

Atomic and atomic ion species have also been detected in a number of transiting planets in the unbound portion of the
atmosphere, or exosphere. The best studied case is HD 209458b where H I, C II, O I, and Si III have been observed
\citep{vidal03,vidal04,linsky10}. Exosphere detections have also been reported in HD 189733b
\citep{lecavelier10,jensen12} and Wasp-12b \citep{fossati10}.

\begin{table*}
\begin{center}
\caption{Full Phase Infrared Photometry of Exoplanets}\label{table_fp_phot}
\begin{tabular}{llll}
\hline\hline Planet & Wavelength & Amplitude (ppm) & Reference \\
\hline 	
CoRoT-1b & 710 nm & 126 $\pm$ 33 &    \citet{snellen07} \\  
$\upsilon$ And b & 24 $\mu$m & 1300 $\pm$ 74 &  \citet{crossfield10} \\
HD 179949  & 8 $\mu$m & 1410 $\pm$ 330 & \citet{cowan07} \\
HD 149026b & 8 $\mu$m & 227 $\pm$ 66  & \citet{knutson09c} \\  
HD 189733b & 3.6 $\mu$m & 1242 $\pm$ 61 &  \citet{knutson12} \\
           & 4.5 $\mu$m &  982 $\pm$ 89 &  \citet{knutson12} \\
           & 8 $\mu$m & $1200 \pm 200 $ &  \citet{knutson07a} \\
HD 209458b  &  4.5 $\mu$m  & $1090 \pm 120$  &  \citet{zellem14} \\
HAT-P-7b & 400 -- 850 nm & 63.7 & \citet{welsh10} \\
WASP-18b & 3.6 $\mu$m & 2960 $\pm$ 90 &   \citet{maxted13} \\
         & 4.5 $\mu$m & 3660 $\pm$ 70 &   \citet{maxted13} \\

\hline\hline
\end{tabular}
\medskip\\
\end{center}
\end{table*}

\subsubsection{Full Phase Photometry}

As well as observations of the transits and eclipses, information on a planet's atmosphere can be obtained from 
observations of the full phase light curve. In the infrared a hot Jupiter will show variations around the cycle 
due to the variation of temperature across its surface. In the optical where reflected light is seen, variations 
will occur due to the change in the illuminated fraction of the disk, as well as due to phase angle dependent 
scattering processes \citep{seager00}. In some cases the light curves are complicated by ellipsoidal variations 
in the star \citep[e.g.][]{welsh10} or the planet \citep[e.g.][]{cowan12b}. Systems with full phase light curves 
at infrared wavelengths showing significant variation around the cycle are listed in table \ref{table_fp_phot}.
In addition full phase light curves due to reflected light are observed in  many of the systems listed in table 
\ref{table_geom_alb}.

Analysis of these light curves has been used to derive maps of the temperature distribution of HD 189733b
\citep{knutson07a} showing a hot spot offset from the substellar point (consistent with models, see section
\ref{sec_mod_3d}). In the case of Kepler-7b, the reflected light phase curve observed by Kepler is interpreted as showing
the presence of patchy clouds \citep{demory13}.

\subsubsection{Polarimetry}

Reflected light from extrasolar planets will be polarized as a result of scattering from cloud and haze particles and 
from molecules. Normal stars are generally found to have very low intrinsic polarizations \citep{kemp87,bailey10}. In 
a hot Jupiter system the polarization in the combined light of the unresolved star and planet is expected to be in the 
range 10$^{-5}$ -- 10$^{-6}$ \citep{seager00}, and will vary around the orbital cycle with the changing phase angle.

While the expected polarizations are small, polarization is a differential measurement that can be made to high 
sensitivity wth ground-based telescopes, and instruments capable of measuring stellar polarization at the one part 
per million level have been developed \citep{hough06,wiktorowicz08}. \citet{lucas09} reported upper limits on the 
polarization of $\tau$ Boo and 55 Cnc in a broad red band (590 -- 920 nm) and set upper limits on the geometric 
albedo of $\tau$ Boo b and 55 Cnc e for Rayleigh scattering models. 

\citet{berdyugina08} reported polarization varying over the orbital cycle of HD 189733b with an amplitude of $\sim 2
\times 10^{-4}$ in the B band. \citet{wiktorowicz09}, however, found no polarization variation in this system with a 99\%
confidence upper limit of 7.9 x 10$^{-5}$ in a 400--675 nm wavelength range. \citet{berdyugina11} then reported further
observations that confirmed a polarization variation, but with a reduced amplitude of 10$^{-4}$ in the U and B bands and
much lower amplitude in the V band. They claim the data is consistent with that of \citet{wiktorowicz09} when the
different wavelengths are taken into account.

While HD 189733b is a system in which polarization might be expected in view of the Rayleigh scattering haze seen in
transmission spectroscopy \citep{pont08,pont13}, the reported polarization amplitudes are too large to be easily
explained. \citet{berdyugina11} report that the polarization is consistent with a Rayleigh-Lambert model with a geometric
albedo of $\sim$0.6 and ``scattered light maximally polarized'' (i.e. 100\%). However, in Rayleigh scattering models a
layer sufficiently optically thick to produce a high geometric albedo has a maximum polarization of only about 30\%
\citep{buenzli09} as a result of depolarization due to multiple scattering. \citet{lucas09} used Monte-Carlo scattering
models to predict a maximum polarization amplitude of 2.6 $\times 10^{-5}$ for HD 189733b.   

\subsection{Atmospheric Structure}

\subsubsection{Inflated Atmospheres}

One result of transit observations is that many hot Jupiters are ``inflated'', with radii significantly larger than
predicted by models \citep{baraffe08,baraffe10}. This inflation is found to be correlated with the level of stellar
irradiation, with inflation becoming apparent for planets receiving incident flux greater than $2 \times 10^8$ erg
s$^{-1}$ cm$^{-2}$ \citep{miller11,demorys11}.

\citet{weiss13} have used data on 138 exoplanets to derive empirical relations between radius, mass and incident flux as
follows:

\begin{equation}
R_P/R_{\earth} = 1.78 (M_P/M_{\earth})^{0.53} (F/\mbox{erg s$^{-1}$ cm$^{-2}$})^{-0.03}
\end{equation}

for $M_P < 150 M_{\earth}$, and

\begin{equation}
R_P/R_{\earth} = 2.45 (M_P/M_{\earth})^{-0.039} (F/\mbox{erg s$^{-1}$ cm$^{-2}$})^{0.094}
\end{equation} 

for $M_P > 150 M_{\earth}$. \\

The reason for this inflation is still debated. \citet{guillot02} showed that the inflated radii could be understood if
$\sim$1\% of the stellar flux received by the planet was transferred into the deep atmosphere below the photosphere. The
observed relationships between inflated radii and incident flux appear consistent with this idea. However, it is unclear
what is the mechanism for transferring energy into the interior. Mechanisms for inflated radii include downward
transport of mechanical energy by atmospheric circulation \citep{showman02}, enhanced opacities
that help to trap heat in the interior \citep{burrows07a}, dissipation of thermal tides \citep{arras10}, and tidal
heating due to a non-zero eccentricity \citep{jackson08,ibgui10}. The Ohmic dissipation model \citep{batygin10, perna10}
uses the interaction of atmospheric winds and the planetary magnetic field to induce electric currents that heats the
interior. \citet{rauscher13} have modelled the process using a 3D model (see section \ref{sec_mod_3d}) and find that
ohmic dissipation can explain the radius of HD 209458b for a planetary magnetic field of 3 -- 10 G. However,
\citet{rogers14} used 3D magnetohydrodynamic simulations of the atmosphere of HD 209458b and found Ohmic dissipation
rates orders of magntiude too small to explain the inflated radius.

\subsubsection{Temperature Structure}

The dayside spectra of hot Jupiters as defined by the Spitzer IRAC colours (table \ref{table_occ} and figure
\ref{fig_dayside}) have been used to derive information on the atmospheric temperature structure. If temperature
decreases with height then the spectrum shows absorption features due to its atmospheric molecules, but a temperature
inversion can cause the same features to appear in emission. A constant temperature (isothermal) atmosphere would shown
no spectral features. 

The presence of a temperature inversion was first suggested in the infrared spectrum of HD 209458b
\citep{knutson08,burrows07a} where a bump in the spectrum at 4.5 and 5.8 $\mu$m can be understood as water vapour in
emission. A number of other cases have been suggested based on Spitzer photometry. It has been suggested that inversions
result from absorption of starlight by an absorber high in the atmosphere. Suggestions for the absorber include TiO and
VO \citep{hubeny03,fortney08} or photochemically produced sulfur compounds \citep{zahnle09}. However, observations have
so far failed to detect the presence of TiO or VO in eclipse or transit spectroscopy in any of these systems.

\citet{knutson10} argue that the presence of an inversion correlates with the activity of the host star, with the
temperature inversions being found for planets orbiting inactive stars, whereas the non-inverted atmospheres occur in
planets orbiting chromospherically active stars.

However, \citet{madhusudhan10} have investigated the degeneracies between thermal inversions and molecular abudnances, and
find it is often possible to fit both inversion and non-inversion models given the limited data points available from
Spitzer photometry.

\subsection{Composition}

\subsubsection{Water vapour, carbon monoxide and methane}

Analogy with brown dwarfs of similar temperatures discussed in section 2 suggests that the most important species in the
near-IR spectra should be H$_2$O, CO and CH$_4$. From the discussion in section \ref{sec_exo_obs} and tables 2 -- 6, it
will be apparent that H$_2$O and CO are indeed detected in quite a number of giant exoplanet systems by a variety of
different methods. Evidence for these molecules is found in spectroscopy of direct imaged planets (section
\ref{sec_direct}), from high resolution cross correlation methods (section \ref{sec_hires} and table \ref{table_hires})
and from secondary eclipse (section \ref{sec_occn}, table \ref{table_occ_spec}) and transit (section \ref{sec_transit}
and table \ref{table_trans}) spectroscopy.

The data on CH$_4$ is less clear. Although it is reported, for example, in the NICMOS transmission spectrum of HD 189733b
\citep{swain08}, high resolution cross correlation studies at the same wavelength do not detect it \citep{dekok13}, but
do detect CO. This suggests a departure from equilibrium chemistry due to vertical mixing as also suggested by
\citet{knutson12} based on Spitzer phase curves.

The spectra of directly imaged planets shown in figure \ref{fig_direct} also show CO, but at best very weak evidence for
CH$_4$. These are all objects that are cool enough to be in the T dwarf range, but actually show spectra more like those
of L dwarfs. The lack of CH$_4$ once again indicates non-equilibrium chemistry
\citep{barman11a,barman11b,skemer13,zahnle14}. Departures from equilibrium chemistry are discussed further in section
\ref{sec_depart_equil}.

Recently, however, CH$_4$ has been detected photometrically in the very cool ($\sim$600K) planetary mass companion GJ
504b \citep{janson13} as described in section \ref{sec_direct}

\subsubsection{Carbon Dioxide}

Up to a few years ago CO$_2$ was not considered to be an important species for exoplanet and brown dwarf atmospheres as
its predicted equilibrium abundance is quite low. Then \citet{swain09a} reported an absorption feature at 2.0 $\mu$m in the NICMOS dayside
emission spectrum of HD189733b that they identified as CO$_2$. This is a relatively weak CO$_2$ band. It has never been
seen in brown dwarfs, for example, whereas the much stronger CO$_2$ band at 4.2 $\mu$m has been seen
\citep{yamamura10,sorahana12}. 

Fitting the NICMOS feature at 2.0 $\mu$m in HD189733b as a CO$_2$ band results in CO$_2$ mole fractions $\sim10^{-3}$
\citep{madhusudhan09,lee12,line12}. This is several thousand times higher than the expected chemical equilibrium
abundance for solar composition \citep{moses11} or the observed CO$_2$ abundances in brown dwarfs \citep{tsuji11}.
Inclusion of non-equilibrium processes such as photochemistry does not substantially increase CO$_2$ abundances
\citep{zahnle09,moses11}. However, CO$_2$ abundances are sensitive to elemental composition increasing quadratically with
increasing metallicity \citep{lodders02,zahnle09}.

The high CO$_2$ abundance is not clearly seen in other observations of HD189733b. In particular the much stronger CO$_2$
bands at 4.2 $\mu$m and 15 $\mu$m are not apparent in the Spitzer secondary eclipse data. Fitting separately to the
NICMOS and Spitzer data, \citet{madhusudhan09} found a much lower CO$_2$ abundance from the Spitzer data consistent with
equilibrium predictions. If a model is required to fit both the Spitzer and NICMOS data simultaneously, as in the
retrieval analysis of \citet*{lee12} the result is a very high CO$_2$ abundance to fit the NICMOS data, and then a
tightly constrained isothermal temperature profile in the upper atmosphere, which hides the strong 4.2 $\mu$m and 15
$\mu$m CO$_2$ bands that would otherwise be present.

An alternative interpreation is to disregard the NICMOS 2 $\mu$m feature, the only evidence pointing to a high CO$_2$
abundance in HD 189733b. \citet{gibson11} have argued that NICMOS observations are too sensitive to the method of
removing systematics to reliably detect molecular species. In that case it is possible to fit the remaining data on the
transmission and dayside emission spectra of HD 189733b very well using equilibrium abundances as shown by
\citet{dixon13} who used the solar composition opacities from \citet{sharp07}. 

\subsubsection{C/O Ratios}

A high C/O ratio was first suggested for the atmosphere of the highly irradiated hot Jupiter WASP-12b
\citep{madhusudhan11} and \citet{madhusudhan12} has suggested that C/O ratio may be an important parameter for
classifying exoplanet atmospheres. If C/O is greater than 1.0 (the solar value is about 0.5) the chemistry changes
substantially for temperatures above about 1500 K, since almost all the oxygen combines with carbon to form CO, and the
abundances of other oxygen bearing species, including H$_2$O and TiO/VO are substantially reduced. The excess carbon also
results in increased abundances of carbon species such as HCN and C$_2$H$_2$. 

Reanalysis of the secondary eclipse data on WASP-12b by \citet{crossfield12b}, with corrections for the effects of a
contaminating star, concluded that the spectrum was well-approximated by a blackbody and that no constraints on its
atmospheric abundances could be set. Other recent studies of the emission and transmission spectra of WASP-12b
\citep{sing13,swain13,mandell13} do not clearly detect any molecular species and do not significantly constrain the C/O
ratio.

\citet{line13b} have investigated the ability to determine C/O ratios using retrieval models (see section
\ref{sec_ret_mod}) and find that with limited data this is very difficult and the retrieved values are biased towards the solar value or a value of
one.  

\subsection{Clouds and Hazes}

The best evidence for cloud or haze\footnote{In using the terms ``cloud'' or ``haze'' here I have followed the
terminology used in the orignal reports. I am not aware of any accepted definition of the difference between these two 
terms, and in this context they likely refer to the same types of particles but haze is usually thinner than cloud and
often occurs at higher altitude.} in giant exoplanets comes from observations of the resolved planets (or planetary mass
objects) HR 8799b, HR 8799c and 2M 1207b where photometry and spectroscopy point to cloudy atmospheres similar to those
of L dwarfs as already discussed in section \ref{sec_direct}. A recent analysis by \citet{skemer13} including
mid-infrared data concluded that patchy clouds as well as non-equilibrium chemistry (to explain the weakness of the 3.3
$\mu$m CH$_4$ band) were needed to fit the data for the HR 8799 planets, whereas a thick cloud model fitted the 2M 1207b
data.

HD 189733b has good evidence for a Rayleigh scattering haze that is visible in both the transmission spectrum observed
during transit \citep{pont13} and in the reflection spectrum from secondary eclipse \citep{evans13}. Rayleigh scattering
(seen as an increase in radius to the blue) is also seen in the transmission spectrum of WASP-12b \citep{sing13} and
WASP-6b \citep{jordan13}.

\citet{demory13} use an analysis of the optical phase curve and secondary eclipse of Kepler-7b to conclude that clouds
must be present and must have an inhomogenous distribution to explain the lack of symmetry in the phase curve. 

The presence of clouds or hazes are suggested in some other systems by essentially featureless transmission spectra that
lack features expected for a clear atmosphere such as Na or H$_2$O absorption \citep[e.g.][]{line13a,gibson13a}.

\section{ATMOSPHERIC MODELS}

\subsection{Types of Models}

Exoplanet and brown dwarf atmospheres occupy a temperature range extending from that of the Solar system planets to
that of the coolest stars. Modelling techniques for these objects can thus adapt techniques both from traditional stellar
atmosphere modelling \citep[e.g.][]{gray05} and those developed for modelling of the Earth and other Solar system planet
atmospheres \citep[e.g.][]{liou02}. These two fields have developed largely independently and have significant difference
in approach that are now becoming apparent, as methods from both fields are applied to the modelling of exoplanet
atmospheres.

 Howvever, the essentials of atmospheric modelling are the same for all such objects. The VSTAR modeeling
code \citep{bailey12}, for example, has been used successfully for objects ranging from terrestrial
\citep{bailey09, cotton12} and giant planets
\citep{kedziora11} in the
Solar system to exoplanets \citep{zhou13,zhou14}, brown dwarfs and cool stars \citep{bailey12}. 

\subsubsection{Stellar atmosphere type models} 

The traditional approach to stellar atmosphere modelling is typified by the ATLAS series of modelling codes
\citep{kurucz70,kurucz93,castelli04}, and the MARCS models \citep{gustaffson08}. Normally with such models the starting
point is an adopted effective temperature $T_{\mbox{\scriptsize eff}}$, surface gravity (usually specified as $\log{g}$
in cgs units), and metallicity [M/H]. Grids of models can then be calculated for different values of these parameters.
The essential stages in such models are: \begin{enumerate} \item Start with an initial estimate for the pressure
temperature structure of the atmosphere specified at a number ($\sim$40--80) of layers. \item For each layer calculate
the composition of the layer. For hotter stars this primarily involves determining the distribution of ionization states
for each element using Saha's equation. For cooler stars some molecules become important and their concentrations are
calculated assuming chemical equilibrium. \item Calculate the opacity (extinction coefficient) of each layer at each
required wavelength taking account of atomic absorption lines, molecular absorption lines and continuum opacity sources
such as bound-free and free-free absorption, collision induced absorptions, Rayleigh and electron scattering. The
wavelength range must cover all wavelengths at which significant energy transport occurs.  \item Solve the radiative
transfer equation to determine the radiative energy flux through each layer. \item Iteratively adjust the temperature
structure of the model, repeating steps 2 -- 4 as required until the model is in energy balance. The total flux through
each layer, including convective energy flux which is normally determined using mixing length theory \citep{henyey65}
must equal $\sigma T_{\mbox{\scriptsize eff}}^4$. \end{enumerate}

The spectrum of the star can then be obtained, either from the last iteration of the model if opacities and radiative
transfer are calculated with sufficient resolution, or from a separate spectral synthesis model.

Models based on essentially this procedure have been developed for brown dwarf and exoplanet atmospheres
\citep[e.g.][]{tsuji96,allard01,barman01,marley02,bsh03}. In order to model the atmospheres of these cooler objects a
number of additional complications have to be dealt with.

At lower temperatures the composition becomes dominated by molecules (as shown in figure \ref{fig2}), and the calculation
of composition (or equation of state) becomes primarily a chemical model. Large numbers of chemical compounds are
potentially important and hence large chemical models handling hundreds or in some case thousands of species have been
developed \citep[e.g][]{lodders02f}. (see section \ref{sec_chem}).

As well as gas phase species, at temperatures below about 2000 K condensates start to form and both modify the gas phase
chemistry, and can form clouds that contribute substantially to the opacity. As we have already seen clouds are important
in understanding the behaviour of L dwarfs and the L/T transition, and are also probably important in giant exoplanets.
(see section \ref{sec_mod_clouds}).

Molecules and cloud particles contribute to scattering of light. Scattering is usually a relatively minor contribution to
the opacity of stellar atmospheres and is usually treated using simplifying approximations such as that of isotropic
scattering. In the cooler atmospheres of brown dwarfs and exoplanets scattering becomes more significant, and more
rigorous treatments of scattering that accurately account for the non-isotropic phase functions may be needed
\citep{dekok11,bailey12}. (see section \ref{sec_radtran}).

\subsubsection{Retrieval Models}
\label{sec_ret_mod}

A different approach to modelling exoplanet atmospheres is shown in a number of recent studies
\citep*{madhusudhan09,line12,lee12,benneke12,line13b,benneke13} that adopt a retrieval approach. These approaches are similar to that
used in remote sensing studies of the Earth atmosphere where temperature structure \citep[e.g.][]{rosenkranz01} trace gas
content \citep[e.g.][]{buchwitz05}, and cloud properties \citep[e.g.][]{garnier12} are routinely retrieved from satellite
observations, and similar techniques are used to study the atmospheres of other Solar system planets from orbiting
spacecraft or Earth-based telescopes.

These models seek to retrieve the temperature structure and composition of the atmosphere directly from observations,
rather than predict these using energy balance and chemical models as in the approach described above. Thus only steps 3
and 4 of the modelling procedure are needed in the forward model. A number of different approaches to the retrieval
process have been used..
\citet{madhusudhan09} search in a large grid of models covering a wide parameter space. \citet*{lee12} use an iterative
optimal estimation procedure. \citet{line13b} investigate several different approaches to retrieval (Optimal Estimation,
Bootstrap Monte Carlo, and Differential Evolution Markov Chain Monte Carlo).

\citet{benneke12,benneke13} use a somewhat different approach where the temperature profile is not retrieved, but is determined by a self
consistent model requiring radiative and hydrostatic equilibrium and allowing for convection. The model can therefore be
described by a small number of parameters (planet-to-star radius ratio, cloud-top pressure and Bond albedo) as well as
the mole fractions of molecular species. The models include cloud and haze layers and retrieval uses a Bayesian Nested
Sampling Monte Carlo method.

\subsubsection{3D Models of Hot Jupiters}
\label{sec_mod_3d}

All the models considered so far are 1D models that describe the structure of the atmosphere with a single one
dimensional profile. Such models cannot represent dynamical effects and diurnal variations.
For hot Jupiters which all receive strong irradiation from their star, the structure is expected to
vary substantially around the planet and can be very different on the dayside and nightside.

\begin{figure*}
\begin{center}
\includegraphics[scale=0.38, angle=0]{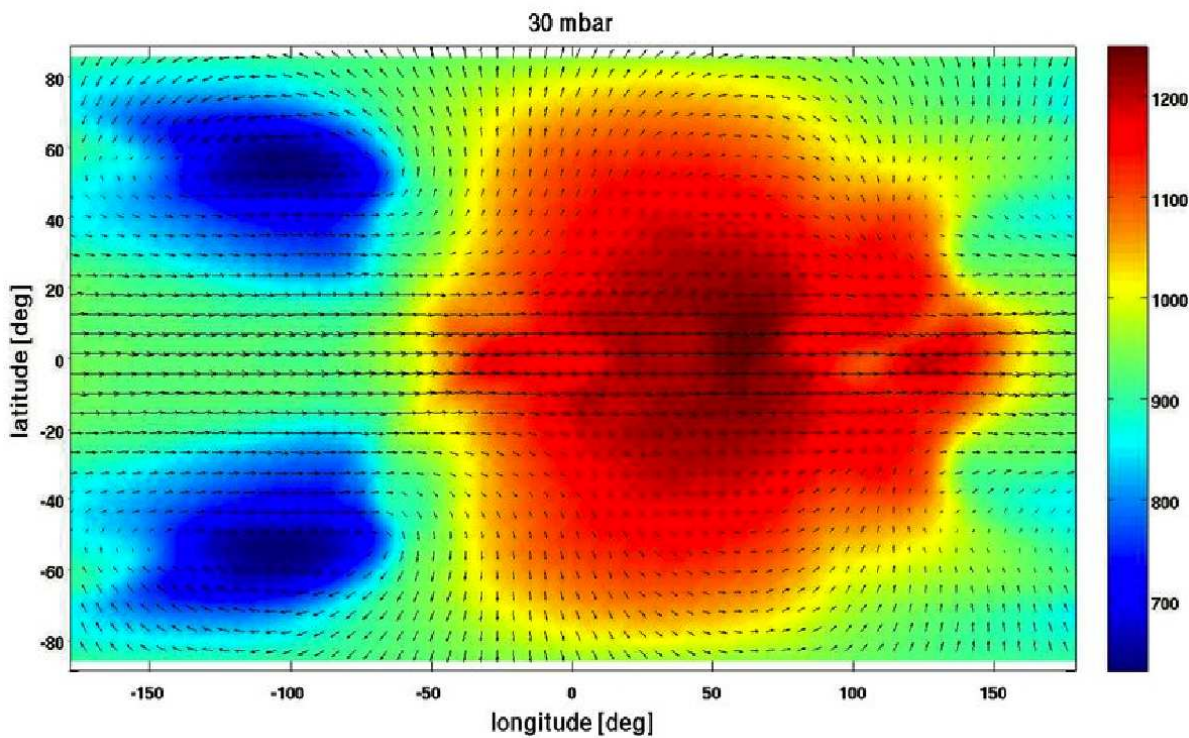}
\caption{Temperature (colourscale in K) and winds (arrows) at the 30 mbar level for a 3D general circulation model 
simulation of the atmosphere of HD 189733b. (Figure 4, Showman, A.P. et al., Atmospheric Circulation of Hot 
Jupiters: Coupled Radiative-Dynamical General Circulation Model Simulations of HD 189733b and HD 209458b, The 
Astrophysical Journal, 699, 564. reproduced by permission of the AAS.)} 
\label{fig_gcm}
\end{center}
\end{figure*}

A number of studies have looked at full 3D models of the atmospheric circulation (General Circulation Models or 
GCMs) for
hot Jupiters. A GCM typically consists of a dynamical core (usually adpated from an Earth atmosphere model) that
numerically solves the equations that govern atmospheric circulation over a three dimensional 
grid of points. These equations can either be the ``primitive equations'' that include the approximations of vertical
hydrostatic equilibrium and a shallow atmosphere as used by \citet{showman09} and \citet{rauscher12} or the full equations that avoid these
approximations as used by \citet{dixon08} and \citet{mayne14}. Initially simplified schemes were used to represent the forcing from the illuminating star
\citep[e.g.][]{showman02,cooper05,menou09} but more recent models include a radiative transfer model and are therfore coupled
radiative-dynamical models. 

Because of the need to perform radiative transfer solutions for each grid point and time step, radiative transfer methods for
GCMs generally need to be simplified compared with those used in the 1D models described earlier. \citet{dixon08}  used a grey
model described by a single mean opacity. \citet{heng11} and \citet{rauscher12} use dual-band radiative transfer, dividing
radiation into incoming shortwave radiation from the star, and outgoing longwave radiation from the planet. \citet{showman09} use
the correlated-k method (see section \ref{sec_spac_abs}) with 30 wavelength bins. \citet{dixon13} use a similar set of 30 bins but
use band-averaged opacities rather than the correlated-k method. The radiative transfer in all these cases uses a two-stream
approximation. \citet{amundsen14} has tested the accuracy of some of these approaches and find that correlated-k and two-stream
methods give reasonable accuracy, but band-averaged opacities can lead to substantial errors.

A review of atmospheric circulation models for
exoplanets is given by \citet{showman10}. Features predicted by such models are the presence of a superrotating jet in the
equatorial regions with wind velocities of 1 -- 4 km/s. This can cause an eastward displacement of the hottest region
from the substellar point which is consistent with observations of HD 189733b \citep{knutson07a,knutson09b}.

Models that specifically aim to simulate the atmospheres of HD 189733b and HD 209458b have been given by
\citet{showman09} and an example of the predicted temperature structure and winds is given in figure \ref{fig_gcm} and
show reasonable agreement with observed day-night phase variations in Spitzer photometry of HD 189733b
\citep{knutson07a,knutson09b}.

\citet{dixon13} have used a 3D model of HD 189733b including wavelength dependent radiative transfer to make predictions
of the transmission spectrum, dayside spectrum, and phase curves that are in good agreement with the observations.

\subsubsection{3D Models of Brown Dwarfs}

Studies of atmospheric circulation in brown dwarfs have been made using 3D models and analytic theory
\citep{showman13} and shallow water models \citep{zhang14}. These show atmospheric circulation with horizontal wind
speeds up to 300 m s$^{-1}$, and vertical mixing that could help to explain the disequilibrium chemistry and patchy
clouds near the L/T transition (see section \ref{sec_bd_var}).

\subsection{Atmospheric Chemistry}
\label{sec_chem}

\subsubsection{Equilibrium Chemistry}

Chemical models for brown dwarf and exoplanet atmospheres aim to predict the chemical composition in the atmosphere given the
pressure, temperature and elemental abundances. Normally this is based on the assumption of chemical equilibrium. This can be
achieved by solving a system of equations for the mass balance of each element and for the overall charge balance using the
equilibrium constants of formation for each compound \citep[e.g.][]{tsuji73,allard01,lodders02f}. An alternative, but equivalent,
approach is that of minimization of the total Gibbs free energy of the system \citep{sharp90,sharp07}.

In either case the required data is available in compilations such as the National Institute for Standards and Technology
(NIST)-JANAF thermochemical tables \citep{chase98} and similar tables such as \citet{barin95} and \citet{robie95}. These tables list
the equilibrium constants of formation $K_f$ and Gibbs free energy of formation $\Delta_fG^o$ for a large number of compounds as a
function of temperature. The two are related through

\begin{equation}
\Delta_fG^o = -RT\ln{K_f}
\end{equation}

Where R is the gas constant. The required thermochemical data for gas phase species can also be derived from spectroscopic
constants.

Chemical models predict the abundances of gas phase species, ionized species and the formation of liquid and solid condensates. The
thermochemical models can also predict quantities such as the mean molecular weight, the specific heat and the adiabatic gradient,
the latter two quantities being needed for mixing length convection theory. 

\subsubsection{Departures from Equilibrium}
\label{sec_depart_equil}

Departures from equlibrium chemistry can occur as a result of photochemistry or vertical mixing if these processes occur at a faster
rate than the collisional processes that tend to restore equilibrium. A non-equilibrium correction to the equilibrium abundances of
CH$_4$/CO and NH$_3$/N$_2$ due to vertical mixing \citep{saumon03} has been adopted to explain the observations of these species in
brown dwarfs. A similar nonequilibrium treatement is used by \citet{barman11a} to model the exoplanet HR8799b. \citet{cooper06} have
found that similar departures from CO/CH$_4$ equilibrium occur in tidally-locked hot-Jupiters.

\citet{zahnle14} have explored the disequilibrium abundances of CH$_4$/CO and NH$_3$/N$_2$ in brown dwarfs and
self-luminous giant planets using a chemical kinetic approach. They find that the low gravity of planets strongly
discriminates against CH$_4$, and that in Jupiter mass planets CH$_4$ becomes more abundant than CO only for temperatures
below about 400 -- 600K depending on the effects of vertical mixing. Ammonia is also sensitive to gravity but insensitive
to mixing making it a potetnial proxy for gravity. 

Chemical models for hot Jupiter atmospheres using a chemical kinetic approach that can include the effects of photochemistry have
been explored in a number of studies \citep[e.g.][]{zahnle09,line10,line11,moses11,venot12,agundez14} \citep[see also
review by][]{moses14}. There remain some differences between model predictions for species such as CH$_4$ and NH$_3$ due to
uncertainties in reaction rates and transport parameters (see discussion in \citealt{moses14} and \citealt{agundez14}), but
generally these models show enhancements of a number of species in the upper atmosphere due to photochemical effects. Most
of these models consider C, N and O containing species.
The model by \citet{zahnle09} includes sulfur species and explores the photochemical production of HS
(mercapto) and S$_2$ as possible absorbers that could contribute to stratospheric heating.

\subsection{Spectral Line Absorption}
\label{sec_spac_abs}

Absorption lines due to rovibrational and electronic transitions of molecules are the most important features of the spectra of
brown dwarfs and planets. Species that are important include H$_2$O, CO, CH$_4$, CO$_2$ and NH$_3$, metal oxides such as TiO and VO,
metal hydrides including FeH, CrH, MgH, CaH and TiH, and carbon species such as CH, CN, C$_2$, HCN and C$_2$H$_2$ (particularly in
carbon rich atmospheres). 

Large numbers of vibrational and rotational levels can be excited at the temperatures of a few thousand degrees encountered in
ultracool dwarfs and hot Jupiters. This leads to a requirement for large line lists containing many millions of lines such as the
BT2 \citep{barber06} computed line list for H$_2$O. 

The spectral line data are used in models to calculate the absorption in each
atmospheric layer. This can be done using on-the-fly line-by-line calculations \citep{allard01,bailey12}
which has the advantage of being the most accurate and flexible method and resulting in high-resolution model spectra.
However it is also the most computationally intensive approach.

A faster approach is to precalcuate opacity tables
\citep{sharp07,freedman08} that are then interpolated for the actual models. However, this can lead to
inaccuracies if the wavelength bins are made too large.  A widely used appoach in
Earth atmosphere modelling is the correlated-k (or k-distribution) method \citep{goody89}, which allows the use of larger 
wavelength bins while retaining accuracy. Recently correlated-k techniques have been used in exoplanet
retrieval models \citep{lee12} and in hot Jupiter GCMs \citep{showman09}.

Sources of spectral line data for the important species have been discussed in detail in a number of recent papers \citep{sharp07,
freedman08, bailey12, tennyson12}. These also discuss related continuum absorption processes and the handling of line shapes. The
reader is referred to these papers for detailed information, and the discussion here relates only to recent developments.

\subsubsection{Carbon Dioxide (CO$_2$)}

The Carbon Dioxide Spectroscopic Databank \citep[CDSD][]{tashkun03} previously available in 296 K and 1000 K versions is now
available in a 4000 K version\footnote{at ftp://ftp.iao.ru/pub} containing lines to an intensity of 10$^{-27}$ cm molecule$^{-1}$ at
4000 K for four isotopologues over the range 226 -- 8310 cm$^{-1}$ (628 million lines).

New computed line lists for CO$_2$ and its isotopologues at 296K and 1000K (the Ames-296K and Ames-1000K lists) have been 
described by \citet{huang13,huang14}.

\subsubsection{Ammonia (NH$_3$)}  

A computed line list for ammonia at temperatures up to 1500 K and containing more than 1.1 billion lines for frequencies up to 12000
cm$^{-1}$ (the BYTe list) is described by \citet{yurchenko12}.

\citet{hargreaves11,hargreaves12a} have provided line lists based on laboratory measurements of NH$_3$ lines at temperatures from
300$^\circ$C to 1400$^\circ$C over the wavelength range from 740 -- 4000 cm$^{-1}$.

\subsubsection{Methane (CH$_4$)}

Methane has been the most problematic of the important species in exoplanet and brown dwarf atmospheres as far as line data is
concerned. Significant recent progress has been made with modelling \citep[e.g.][]{rey13,nikitin13b,yurchenko13}  and a large
computed line list for hot methane has very recently been developed \citep{yurchenko14}\footnote{available at
http://www.exomol.com/data/molecules/CH4/12C-1H4}. \citet{yurchenko14b} have shown that using this line list it is possible to
obtain good model fits to the methane bands in the near infrared spectra of brown dwarfs that could not be fitted with older line
lists, such as those based on the Spherical Top Data System software \citep[STDS][]{wenger98}. Another computed line list for
hot methane has been reported by \citet{rey14} but is limited to wavelengths longer than 2 $\mu$m.

Much improved line lists for the 1.26 -- 1.71 $\mu$m region at temperatures from 80 -- 300 K have been developed recently from
extensive laboratory measurements at cryogenic and room temperature \citep{wang12,campargue12a,campargue13}. These lists, and
earlier versions of them, have been used successfully for modelling the spectra of Titan \citep{bailey11,debergh12,campargue12b} and
Uranus \citep{irwin12,bott13}. An improved low temperature line list has also been developed for the 2 $\mu$m region
\citep{daumont13}. These lists have been incorporated into the new 2012 edition of the HITRAN
database\footnote{www.cfa.harvard.edu/hitran} recently released.

Empirical line lists for methane measured at temperatures from 300 -- 1400 $^\circ$C over the wavelength range from 2.0 -- 10.4
$\mu$m have been provided by \citet{hargreaves12b}.

\subsubsection{SiO and HCN/HNC}

New line lists for SiO \citep{barton13} and HCN/HNC \citep{barber14} have recently been published by the ExoMol group.

\subsubsection{Collision Induced Absorptions (CIA)}

The collision induced absorption of H$_2$ - H$_2$ and H$_2$ - He pairs are important contributors to the opacity of brown dwarfs and planets. Updated data on these absorptions have recently been provided by \citet{abel11,abel12} as described in \citet{saumon12}. This data as well as other CIA
datasets have been recently made available in a new section of the HITRAN database \citep{richard12}.

\subsection{Condensates and Clouds}
\label{sec_mod_clouds}

Condensed phases (i.e. solids and liquids) begin to condense out of the gas when temperatures drop to around 2000 K and lower. These
condensates can form clouds that can significantly alter the opacity and hence the structure of the atmosphere. Chemical models
(section \ref{sec_chem}) can predict which species will condense (these include oxides, silicates and iron) and the amounts of
condensed material produced. However, it is harder to predict what size particles will be produced and whether they will remain in
place as clouds or fall under gravitation (precipitation, sedimentation or rain-out).

Lorenz-Mie scattering theory can be used to predict the optical properties of the cloud particles. In the general case
these include the extinction coefficient, the single scattering albedo (the fraction of light that is scattered
rather than absorbed) and the phase function that describes the angular distribution of scattered light. These
are needed as inputs for solving the radiative transfer equation (equations \ref{eqn_rt} and \ref{eqn_source},
section \ref{sec_radtran}).

\subsubsection{Clouds in Brown Dwarfs}

Two limiting cases were considered in the COND and DUSTY models of \citet{allard01}. The COND models include condensate formation,
which alters the chemistry by depleting elements from the gas, but did not include any contribution of the condensates to the
opacity. In the DUSTY models the condensed material is assumed to remain in place in equilibrium with the gas phase and form clouds
of small dust grains. The DUSTY models were found to be a good representation of late-M and early-L dwarfs, but at cooler
temperatures they produce weakening of spectral features and increasingly red colours in disagreement with the observations of L-T
transition objects.

\begin{figure}
\begin{center}
\includegraphics[scale=0.43, angle=0]{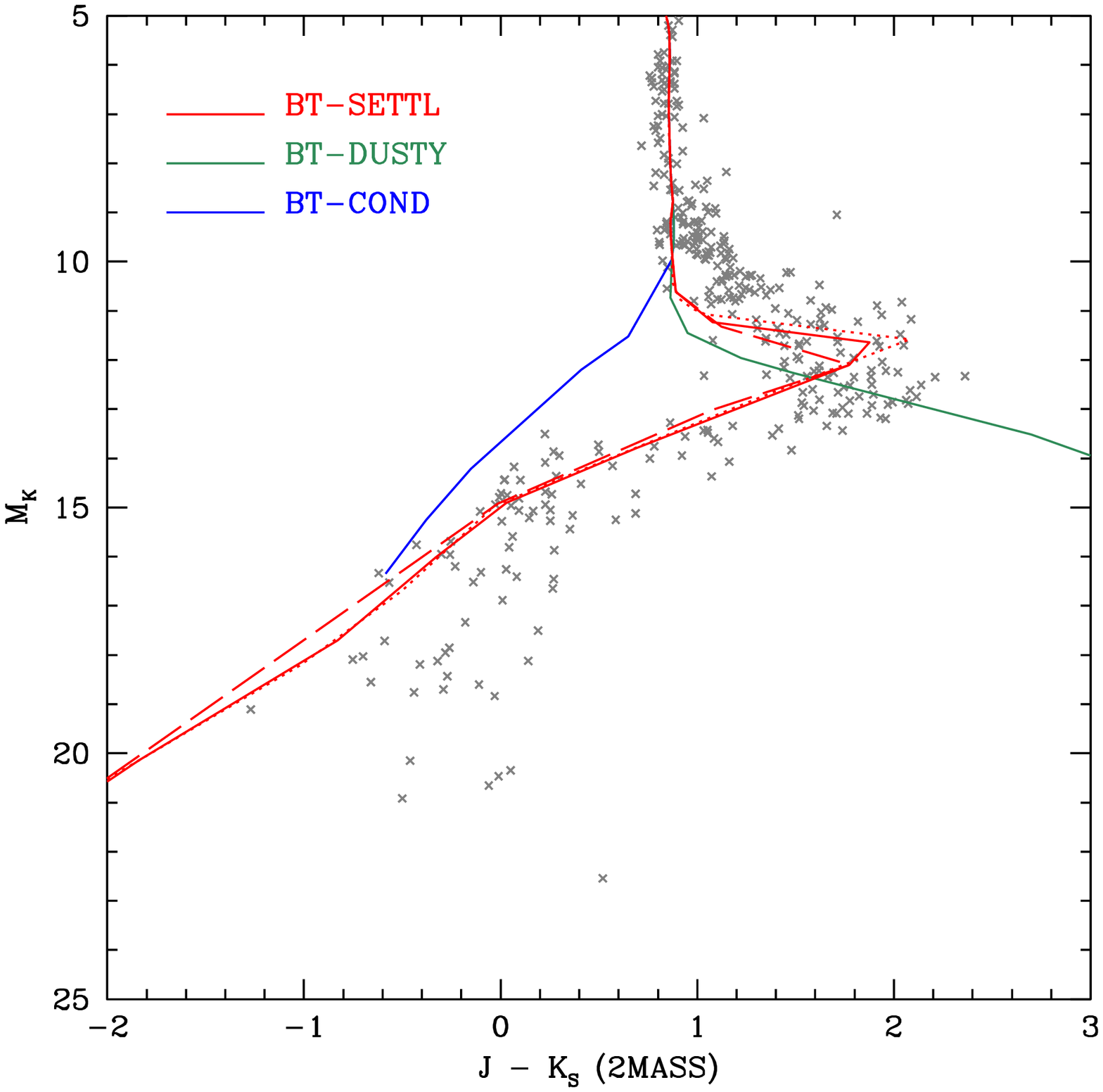}
\caption{Colour magintude diagram using the same data as figure \ref{fig_phot} compared with the predictions of model atmospheres using different cloud models. The BT-COND and BT-DUSTY models are updated version of the COND and DUSTY models of \citet{allard01} with more modern opacities. The BT-SETTL model is described by \citet{allard07,allard12}. The plotted lines are predicted synthetic magnitudes for the isochromes of \citet{baraffe03} and \citet{chabrier00a}. For BT-SETTL 1, 3 and 5 Gyr isochrones are plotted as dotted, solid and dashed lines. For COND and DUSTY the 3 Gyr isochrone only is plotted. } 
\label{fig_phot_mod}
\end{center}
\end{figure}

The cloud-free COND models were found to be a fairly good representation of mid to late T dwarfs, indicating that gravitational
settling has largely removed dust from the atmospheres in these cases. However, neither of these two models could account for the
late-L to early T dwarfs. A number of cloud models have now been developed that aim to reproduce the behaviour of clouds through the
full brown dwarf spectral sequence.

In the Unified Cloudy Model \citep{tsuji02,tsuji05}, clouds are assumed to be restricted to a small range of temperatures between
the condensation temperature $T_{cond}$ and a critical temperature $T_{cr}$. Below the critical temperature it is assumed that
grains will grow to such a size that they will rapidly precipitate under gravity. A fixed particle size (r = 0.01$\mu$m) is used in
the clouds. The critical temperature $T_{cr}$ is an adjustable parameter, with values in the range 1700 -- 1900 K providing a
reasonable match to the observations.

\citet{burrows06} describe a cloud model that similarly restricts the cloud extent but includes an exponential decay in cloud
particle density at the upper and lower edges of the cloud. They investigate the effects of various cloud parameters and conclude
that cloud particle sizes of 50 -- 100 $\mu$m fit the data best. This is much larger than the grain sizes used in most other models
which are around 1$\mu$m or smaller.

\citet{ackerman01} describe a cloud model based on a balance between turbulent diffusion and sedimentation in horizontally uniform
cloud decks. The model involves a scaling factor $f_{sed}$ that describes the efficiency of sedimentation and typically ranges from
1 to 5. Small $f_{sed}$ values produce thicker clouds and match observations of L dwarfs and higher values are found for later type
T dwarfs \citep{stephens09}.

The BT-Settl models \citep{allard07,allard12} use a cloud treatment based on a model for cloud microphysics from analysis of solar
system atmospheres \citep{rossow78} that predicts timescales for condensation, sedimentation and coagulation. These are compared
with the turbulent mixing timescale to predict grain densities and sizes.

\citet{woitke03,woitke04} and \citet{helling06} have developed a kinetic (non-equilibrium) model for the nucleation, accretion,
gravitational settling and evaporation of dust grains. A version of this cloud model has been integrated with the PHOENIX stellar
atmosphere code \citep{hauschildt99} to provide the DRIFT-PHOENIX models for substellar atmospheres \citep{helling08a}.

A more detailed description of some of these different cloud models and a comparison of their predictions in test cases can be found
in \citet{helling08b}.

A specific aim of these models is to explain the changes that occur in brown dwarfs at the L/T transition as discussed in section 2.
Figure \ref{fig_phot_mod} shows that the BT-Settl model (and other cloud models make simiar predictions) can explain the general
trend seen in the near-IR colour magnitude diagram of a swing from red to blue colours at the L/T transition. The models achieve
this mostly because the cloud has a limited extent in temperature, and so for cooler models the clouds drop to layers below the
photosphere where the effect on the spectra and colours become small.

However, all current models fail to match the details of the L/T transition. As can be seen in figure \ref{fig_lt_trans} models fail
to reproduce the sharpness of the transition as a function of effective temperature. Models also fail to reproduce the J-band
brightening (see section \ref{sec_bd_phot}). The BT-Settl model also predicts J$-$K colours that continue to get bluer with lower
effective temperatures, while observations show fairly constant J$-$K for mid to late T dwarfs (figures \ref{fig_lt_trans} and
\ref{fig_phot_mod}).

Cloud species that condense at lower temperatures (including Cr, MnS, Na$_2$S, ZnS and KCl) are considered by \citet{morley12}, and
found to be helpful in explaining the colours and spectra of late-T and Y dwarfs \citep{leggett13}.

\subsection{Radiative Transfer}
\label{sec_radtran}

Atmospheric models can differ significantly in their handling of radiative transfer, particularly in regards to the treatment of
scattering. Radiative transfer involves the flow of radiation through an atmosphere as determined by the processes of absorption,
emission and scattering. The radiative transfer equation can be written as \citep{bailey12}:

\begin{equation}
\mu \frac{dI_{\nu}(\tau,\mu,\phi)}{d \tau} = I_{\nu} (\tau, \mu, \phi) - S_{\nu}
(\tau, \mu, \phi)    \label{eqn_rt}
\end{equation}

\noindent where $I_{\nu}$ is the monochromatic radiance (sometimes referred to as intensity or specific
intensity) at frequency $\nu$, and is a function
of optical depth $\tau$, and direction $\mu$, $\phi$, where $\mu$ is the cosine of
the zenith angle, and $\phi$ is the azimuthal angle. The source function $S_{\nu}$ is
given by:

\begin{align}
S_{\nu} (\tau,\mu,\phi) & = \frac{\varpi(\tau)}{4 \pi} \! \! \int_0^{2 \pi} \! \! \! \!
\int_{-1}^{1} \! \! \! \! P(\mu,\phi; \mu', \phi') I_{\nu} (\tau, \mu', \phi') d \mu' d \phi' 
\nonumber \\
  &  \mbox{} + (1 - \varpi) B_{\nu} (T)  \label{eqn_source} \\
  &  \mbox{} + \frac{\varpi F_{\nu}}{4 \pi} P(\mu,\phi; \mu_0, \phi_0) \exp{(-\tau / \mu_0)}
  \nonumber
\end{align}

\noindent where the first term describes scattering of radiation into the beam from other
directions according to single scattering albedo $\varpi$ and phase function $P(\mu,\phi; \mu', \phi')$, the second term
is thermal emission, with $B_{\nu} (T)$ being the Plank function and the third term is direct illumination of the atmosphere by
an external source with flux $\mu_0 F_{\nu}$ and direction $\mu_0, \phi_0$ (e.g. the Sun or host star).

It is the first term in equation \ref{eqn_source} involving the double integral that significantly complicates radiative transfer.
This term has the consequence that the radiance in any one direction is dependent on the radiance in all other directions (since any
of these can potentially scatter into the line of sight). In general it is then only possible to solve for the full angular
dependence of the radiation field in all directions.

To avoid this complication the handling of scattering is often simplified, in some cases by ignoring it entirely, or by using a
simplified form for the phase function $P$ such as the assumption of isotropic scattering, and/or a simplified form for the angular
depedence of $I_{\nu}$ such as the two-stream approximation or the Eddington approximation. In stellar atmospheres approximate
methods can be justified by the fact that scattering is generally of minor importance, and where it does become significant, in the
form of Rayleigh scattering from molecules in cool stars, the phase functions are at least forward-backward symmetric.

Where clouds are present, however, the phase functions can be highly non-isotropic, and in the case of Solar system planet
atmospheres, radiative transfer methods that more rigorously handle multiple scattering with non-isotropic phase functions are
generally used. These include, in particular, versions of the discrete ordinate method originally due to \citet{chandra60} which has
been developed into robust and general radiative transfer solving codes such as DISORT \citep{stamnes88}, SHDOM \citep{evans98} and
LIDORT \citep{spurr01}. DISORT is used by \citet{bailey12} in the VSTAR code to model brown dwarf spectra. Another appropriate
method is the adding-doubling method \citep{dekok11}.

At present such methods are rarely used in exoplanet and brown dwarf atmospheric modelling, and this opens up the possibility of
errors being introduced due to an oversimplified treatment of scattering. This was investgated by \citet{dekok11} for the thermal
emission spectra of exoplanets who found that substantial errors can be introduced by neglecting scattering, or by using an
isotropic scattering approximation where large particles are present.

\subsection{Polarization}

Scattering processes polarize light, so a full treatment of radiative transfer should take account of polarization. Light scattered
from planetary atmospheres is expected to be polarized whereas the light of normal stars is mostly unpolarized \citep{bailey10}, and
this polarization can potentially be used as a means of differentially detecting planets in imaging observations
\citep{schmid05,keller06}, and as a means of characterising extrasolar planet atmospheres by observing the phase variation of
polarization \citep{seager00,bailey07}. Polarization has also been measured in some brown dwarfs \citep{menard02,zapatero05,tata09}
and is thought to result from scattering in the dust clouds combined with either rotational oblateness or a patchy cloud structure
\citep{sengupta10}.

Polarization is particularly useful for determining the nature and size of cloud particles. A classic application of polarization
was the analysis of the polarization phase curve of Venus by \citet{hansen74} to determine that the clouds of Venus were composed of
$\sim$1 $\mu$m radius sulfuric acid droplets.

Polarization should also, ideally, be included in all radiative transfer modelling involving scattering, because even if we are not
interested in observing polarization, neglecting polarization can alter the resulting fluxes. \citet{stam05} investigated this for
reflected light from extrasolar giant planets and found that errors in geometric albedo of up to 9\% could arise as a result of
neglecting polarization. In practice, however, polarization is normally ignored in radiative transfer calculations, because
including polarization would substantially slow down the computations.

Polarization in Earth-like planet atmospheres will be discussed later in section \ref{sec_sig_hab}.

\section{THE SEARCH FOR HABITABLE PLANETS AND LIFE}

The main focus of this review has been on the study of planetary atmospheres for which we have observations, and so far this has
been almost entirely giant planets. The only exceptions to this are the super-Earths GJ 1214b and HD 97658b. Transit
spectroscopy of GJ 1214b has been
obtained \citep{bean10,crossfield11,berta12,kreidberg14a} showing a featureless spectrum indicating an atmosphere either rich in
water vapour, or with high altitude clouds. Transit spectroscopy of HD 97658b \citep{knutson14b} also shows a featureless spectrum
inconsistent with a cloud-free solar metallicity atmosphere.

In this section the potential for characterization of Earth-like planets in the habitable zone of their stars is briefly reviewed.

\subsection{The Habitable Zone}

The habitable zone is defined as the range of distances from its star at which a planet has suitable conditions for liquid water to
be able to exist at its surface. In the absence of an atmosphere the average surface temperature $T_{eq}$ of a planet is given by
energy balance between radiation received from its star, and thermal radiation to space as:

\begin{equation}
 (1-a)S/4 = \sigma T_{eq}^4
\end{equation}

Where $S$ is the total incident flux (W m$^{-2}$) received from the star (in the case of the Earth this is the solar constant S$_0$
= 1361 W m$^{-2}$), $a$ is the Bond albedo of the planet and $\sigma$ is the Stefan-Boltzmann constant. The factor of 4 comes from
the fact that radiation received over an area of $\pi r^2$ is redistributed over the entire surface of the planet with area $4 \pi
r^2$. For Earth this calculation gives an equilibrium temperature of $T_{eq} \sim 255$ K currently, and lower in the past as the
solar luminosity increases with time and was about 30\% less early in the Sun's history \citep{bahcall01}. 

The global average temperature of the Earth is, of course, higher than this at about $T \sim 288$ K, with the difference being due
to the operation of the greenhouse effect that traps some of the outgoing radiation and causes the outgoing flux to be less than
$\sigma T_{eq}^4$. In general, from observations of the orbit of a planet we can determine $S$, but in most cases we won't know the
albedo $a$ or the amount of the greenhouse effect, and so can't directly determine the surface temperature of a planet from
observations. According to \citet{selsis07} $T_{eq}$ must be less than 270 K for a planet to be habitable.

Estimates of the locations of the edges of the habitable zone have been made based on the assumption of an Earth-like planet with a
CO$_2$/H$_2$O/N$_2$ atmosphere using cloud-free 1D radiative-convective models. \citet{kasting93} gave the extent of the habitable zone from 0.95 AU to 1.37 AU for our Solar
system, with the inner edge being set by the onset of the moist greenhouse process \citep{kasting88} causing loss of water to space,
and the outer edge being set by cooling due to the formation of CO$_2$ clouds. However \citet{forget97} showed that CO$_2$ clouds
actually cause warming and allow a more extended habitable zone. An upadated calculation is given by \citet{kopparapu13} which sets
the moist greenhouse inner edge at 0.99 AU, and the outer edge at 1.67 AU based on the maximum greenhouse criterion. On this basis
the Earth is near the inner edge of the habitable zone.

These results can be scaled for other stars according to $S/S_0$, the flux received by the star as a fraction of the solar constant
and the effective temperature of the star. A habitable zone calculator for this puropse based on the results of \citet{kopparapu13}
is available\footnote{http://depts.washington.edu/naivpl/content/hz-calculator}. The effects of different planet masses on the
position of the habitable zone are considered by \citet{kopparapu14} who find that the inner edge moves so as to give a wider
zone for higher mass planets..

Recent studies using 3D climate models, however, have found the inner edge of the habitable zone at $\sim$0.95 AU \citep{leconte13}
or $\sim$0.93 AU \citep{wolf14}, significantly smaller than the 1D models described above.

The results assume an Earth-like planet and could be different for other types of planets. \citet{abe11} have shown that a more
extended habitable zone is possible for a desert planet with limited surface water, and \citet{zsom13} find a miminum inner edge for
the habitable zone of 0.38 AU for hot desert worlds. \citet{pierrehumbert11} have suggested that the greenhouse effect due to
collision induced absorption in molecular hydrogen could allow habitable conditions to be maintained out to 10 AU from a G-type
star.

\subsection{Habitable Zone Planets}

\begin{table*}
\begin{center}
\caption{Low Mass Planets in or near the Habitable Zone} \label{tab_hz_planets} 
\begin{tabular}{llllllll}
\hline Planet & \multicolumn{2}{c}{Star} & Period & M $\sin{i}$ & Radius & $S/S_0$ & Reference \\
       & Type & T$_{\mbox{\scriptsize eff}}$ & (days) & M$_{\earth}$ & R$_{\earth}$ & & \\
\hline\hline 
$\tau$ Cet e & G8.5V & 5344 & 168.1 & 4.3 &  & 1.60 & \citet{tuomi12} \\
HD 40307g & K2.5V & 4956 & 197.8 & 7.1 & &  0.62 & \citet{tuomi13} \\  
HD 88512b & K5V & 4715 & 58.43 & 3.6 &  & 1.86 & \citet{pepe11} \\
GJ 163c  & M3.5  &      & 25.63 & 6.8 &  & 1.34 & \citet{bonfils13} \\
GJ 667C c & M1.5V & 3350 & 28.14 & 3.8 &  & 0.90 & \citet{anglada12,anglada13} \\
GJ 667C e & M1.5V & 3350 & 62.24 & 2.7 &  & 0.33 & \citet{anglada13} \\
GJ 667C f & M1.5V & 3350 & 39.03 & 2.7 &  & 0.58 & \citet{anglada13} \\
GJ 832c   &  M1.5V  &  3472 &  35.68 &  5.4 &  &  0.87 &  \citet{wittenmyer14} \\
Kapteyn b  &  M1.0  &  3570 &  48.616 &  4.8 &  &  0.4 & \citet{anglada14}  \\
Kepler-22b & G5V & 5518   & 289.9 & & 2.38 & 1.09 &  \citet{borucki12} \\
Kepler-61b & M0 & 4017  & 59.88 &  & 2.15 & 1.26  & \citet{ballard13} \\
Kepler-62e & K2V & 4925 & 122.4 & & 1.61 & 1.2 $\pm$ 0.2 & \citet{borucki13} \\
Kepler-62f & K2V & 4925 & 267.3 & & 1.41 & 0.41 $\pm$ 0.05 & \citet{borucki13} \\
Kepler-69c & G4V & 5638 & 242.5 & & 1.71 & 1.91  & \citet{barclay13,kane13} \\
Kepler-186f &  M1V &  3788 &  129.9 & &  1.11 &  0.32$_{-0.04}^{+0.06}$ & \citet{quintana14} \\
\hline
\end{tabular}
\end{center}
\end{table*}

Table \ref{tab_hz_planets} lists planets that have been reported as being in or near the habitable zone with M $\sin{i} <
10M_{\earth}$ or $R < 2.5R_{\earth}$. Note that the planet of $\tau$ Cet is only reported as a tentative detection \citep{tuomi12}
and the reality of some of the planets of GJ 667C have been disputed \citep{gregory12,feroz14}. The reported
habitable zone planets of GJ 581 \citep{mayor09, vogt10} have been excluded from the table based on the analysis
of \citet{robertson14}.

\citet{petigura13} have analysed Kepler data to find 10 planet candidates with radii of 1 -- 2 R$_{\earth}$ and within a habitable
zone defined by 0.25 $< S/S_0 <$ 4. Allowing for incompleteness they estimate that Earth-size planets in the habitable zone occur in
22 $\pm$ 8 \% of stars. With the narrower definition of the habitable zone discussed above \citep[0.99 -- 1.67 AU, ][]{kopparapu13}
this becomes 8.6 \%. 

\subsection{Detecting and Characterizing Earth-Like Planets}
\label{sec_det_char}

The direct detection and characterization of Earth-like planets is far more challenging than for the giant planets discussed in
section \ref{sec_giant}. The contrast ratio between an Earth-like planet and its star is $\sim10^{-10}$ at visible wavelengths and
$\sim10^{-7}$ in the thermal IR ($\sim10 \mu$m). 

One concept is that of an infrared interferometer in space as first suggested by \citet{bracewell78}. This was developed into the
Darwin \citep{cockell09} and Terrestrial Planet Finder Interferometer \citep[TPF-I, ][]{beichmann99} mission concepts. These
involved several infrared telescopes flying in formation in space and combining their light to achieve nulling interferometry, so
that the light of the star could be suppressed, and reveal the light of the planet. These missions would have aimed to both detect
planets and obtain low resolution spectroscopy over the 6 -- 20 $\mu$m range for atmospheric characterization and biosignature
detection. 

An alternative concept was the Terrestrial Planet Finder Coronograph \citep[TPF-C, ][]{traub06}. This was envisaged as a space
telescope with an 8 by 3.5 m elliptical mirror, using advanced coronographic techniques to suppress starlight at the 10$^{-10}$
level. It operated at visible wavelengths and would be able to detect planets and carry out spectroscopic characterization. 

Both Terrestrial Planet Finder missions (TPF-I and TPF-C) have now been cancelled by NASA, and the ESA Darwin mission study ended in
2007.

A further concept for starlight suppression involves the use of an occulter (or starshade) placed in front of the telescope. The
occulter must use a petal shaped design to suppress diffraction and be placed about 40,000 km in front of the telescope
\citep{cash06}. An occulter could be used in conjunction with the James Webb Space Telescope \citep{soummer09} and/or with a
dedicated space telescope as in the New Worlds Observer \citep[NWO, ][]{turnbull12} mission concept. NWO would use a 4m telescope
and enable detection and spectroscopic characterization of exoplanets with R $\sim$ 100 over 0.3 -- 1.6 $\mu$m. The mission aims to
achieve a 95\% probability of detecting and characterising at least one habitable zone Earth-like planet. A more recent
starshade mission concept is the Exo-S mission described in section \ref{sec_future}.

Extreme adaptive optics systems on giant ground-based telescopes are another potential approach to the detection and
characterization of Earth-like exoplanets. However, a number of studies have concluded that such systems on currently planned
extremely large telescopes fall well short of the required sensitivity \citep{stapelfeldt05,mountain09}. However, \citep{angel03}
has suggested that detection with a 20m ground-based telescope and spectroscopic characterization with a 100m telescope should be
feasible taking advantage of an Antarctic site. 

\citet{kaltenegger09} considered the feasibility of characterising Earth-like planets using transit spectroscopy from a 6.5m
telescope in space (e.g. the JWST). They found that the signal-to-noise values for all important spectral features were of the order
of unity or less per transit.

The situation for transit observations is much improved if the Earth-like planet is assumed to be in the habitable zone of an
M-dwarf rather than a solar type star. This leads to both a larger transit signal, since the star is smaller, and more frequent
transits. \citet{palle11} conclude that detection of atmospheric features in transiting Earth-like planets could be possible in
planets orbiting very cool stars or brown dwarfs with the proposed 42 m European Extremely Large Telescope.

\citet{snellen13} have suggested the use of high resolution cross correlation techniques (see section \ref{sec_hires}) to detect
oxygen absorption during the transit of an Earth-like planet across a red dwarf star. They suggest this should be feasible with a
ground-based telescope with an effective aperture of $\sim$100 m and suggest this could be an array of low-cost ``flux collectors''
which would not require high image quality.

\subsection{Signatures of Habitability}
\label{sec_sig_hab}

The presence of a planet within the habitable zone does not necessarily mean that it has habitable conditions on its surface. The
best signature of habitability would be direct detection of the presence of liquid water. This is difficult to achieve using
spectroscopy. While atmospheric water vapour can be detected through absorptions in the near-IR or thermal infrared, the presence of
surface liquid water does not provide any clear spectroscopic signature.

A possible indicator of liquid water would be the presence of the ``glint'' signal due to specular reflection from oceans.
\citet{williams08} modelled the light curves and polarization phase depenence for Earth-like planets and showed that distinctive
signals due to glint should be detectable for planets with surface oceans. \citet{robinson10} used an Earth simulation to show that
glint increases the brigthness of the Earth by as much as 100\% at crescent phases. \citet{zugger10,zugger11} modelled glint effects
including polarization and found that the glint signal becomes diluted and more difficult to detect when clouds, aerosols and
surface winds are included. \citet{cowan12a} show, however, that latitude dependent albedo variations produce a signal in the phase
curve very similar to that from glint, and therefore the glint signal may not be a reliable indicator of the presence of oceans.

Another potential way of detecting the presence of oceans is to use rotational changes in the brightness and colours
\citep{ford01,kawahara10}. Such observations can in principle determine the fraction of ocean and land coverage and even provide
maps of the distribution. These ideas have been tested using observations of the integrated Earth from the EPOXI mission with the
Deep Impact spacecraft \citep{cowan09,fujii11,cowan13}. \citet{langford09} used lunar Earthshine measurements to detect photometric
changes associated with the passage of the specular reflection glint spot from land to ocean.

Liquid water clouds in a planet's atmosphere could be detected through the presence of the primary rainbow which would appear as a
peak in the phase curve at a phase angle of about 40 degrees. While the rainbow peak could be visible in the phase light curve, it
is better defined, particularly for small cloud particles, in the polarization phase curve \citep{bailey07,karalidi12}. The size of
the disk integrated rainbow polarization signal for the Earth has not been measured, but in principle could be obtained from lunar
earthshine polarization measurements. Current data however does not have sufficient phase coverage
\citep{sterzik12,takahashi13,bazzon13}. 

\subsection{Biosignatures}

The ultimate aim of such studies is to make observations that would test for the presence of life on an exoplanet. A number of
potential biosignatures have been suggested \citep[e.g. ][]{desmarais02} and tests have been made to determine whether these are
actually visible in observations of the integrated Earth using spacecraft \citep{livengood11} or lunar Earthshine measurements
\citep{woolf02}. Models of the integrated Earth spectrum have been presented by \citet{tinetti06} and \citet{robinson11}.

The most important biosignature is generally considered to be atmospheric oxygen (O$_2$) or ozone (O$_3$). Oxygen is produced in the
Earth atmosphere primarily by photosynthetic organisms. Possible abiotic sources of atmospheric oxygen that could lead to ``false
positivies'' have been discussed \citep{schindler00,selsis02,segura07} but such cases appear to be unlikely, or can be excluded on
the basis of other observations. Atmospheric O$_2$ is detectable through a number of electronic absorption bands with the strongest
being the A-band at around 760 nm. The longest wavelength O$_2$ band with significant strength is the a-X band at 1.27 $\mu$m. The
lack of O$_2$ bands further into the infrared is a problem for life detection with instruments that work optimally at near-IR
wavelengths (e.g. ground-based telescopes with extreme adaptive optics systems that work best at longer wavelenths). However, in the
thermal infrared the band of ozone (O$_3$) at 9.7 $\mu$m can be used. This is also considered to be a good biosignature since O$_3$
is a photolytic by-product of O$_2$.

Methane (CH$_4$) is another potential biosignature. On Earth it originates largely from biological processes (methanogenic archaea),
but there are also possible abiotic sources such as serpentenization. The simultaneous presence of both oxygen and methane was
suggested to be a good biosignature by \citet{lovelock65} indicating chemical disequilibrium. Methane has strong absorption bands at
7 and 3.3 $\mu$m and a series of weaker bands through the near-IR. \citet{segura05} have suggested that CH$_4$ and other reduced
biogenic gases such as N$_2$O and CH$_3$Cl might be useful as biosignatures in planets around M dwarfs where these gases would have
longer photochemical lifetimes than on Earth. 

Another possible biosignature is the ``red edge'', the sharp edge in the reflectance spectrum from vegetation at around 700nm. This
is a strong signal in light reflected directly from a vegetated area. However, in the integrated Earth, lunar Earthshine
observations show a maximum effect of a few per cent \citep{hamdani06,arnold08}. \citet{kiang07} suggested that spectral signatures
could be different for photosynthetic pigments adapted for different stellar types, and \citet{sanroma14} investigate the spectra of
the Archaean Earth when purple bacteria were widespread giving rise to a slightly longer wavelength signal.

\section{The Future}
\label{sec_future}

The characterization of exoplanet atmospheres has made substantial progress over the last few years and some aspects of their
composition and structure are beginning to be resolved. However, there are still many uncertainties and controversies that remain.
The major limitation is in the observational data which are, in most cases, extremely limited in spectral resolution and wavelength
coverage. Data from space instruments have proved most valuable for studying the transmission and dayside emission spectra of
transiting planets, and so the next major advance is likely to be the James Webb Space Telescope due for launch in 2018. Its NIRSpec
and MIRI instruments should make possible spectra of transiting planets across the near-IR and mid-IR spectral regions
\citep{shabram11,belu11}.

Another important mission is TESS \citep{ricker14} that will carry out an all sky survey of bright stars for transiting planets.
While it is not directly aimed at atmospheric characterization, it may well find some of the best targets for more detailed studies.
The ESA PLATO mission planned for 2024 launch \citep{rauer13} is another mission that will search for planets transiting bright
stars, and will target up to 1,000,000 stars. An exoplanet characterization mission EChO \citep{tinetti12} was also proposed for the
same launch opportunitiy but was not selected.

NASA is currently studying three missions with relevance to exoplanet characterization. The Wide-Field Infrared Survey
Telescope - Astrophysics Focused Telescope Assets (WFIRST-AFTA) concept \citep{spergel13} is a 2.4m wide field near infrared (0.6 --
2.0 $\mu$m) telescope. It is mainly aimed at wide field surveys, but the design includes a coronograph that enables it do
characterization observations of exoplanets by direct imaging. Exo-C \citep{stapelfeldt14} is a 1.5m unobscured Cassegrain 
telescope with a coronograph providing imaging of exoplanets over the wavelength range 450 -- 1000 nm. Exo-S \citep{seager14} is a
1.1m telescope using the starshade concept (see section \ref{sec_det_char}) to provide imaging over the 400 -- 1000 nm range. Exo-S should
have the capability to detect Earth-size planets in the habitable zones of about 20 Sun-like stars.

New facilities on ground-based telescopes include improved instruments for direct imaging such as the Gemini Planet Imager
\citep[GPI][]{macintosh12} and SPHERE for the VLT \citep{bezuit10}. The new generation of extremely large telescopes now under
development will open up new possibilities with planned instrument such as EPICS for the 42m E-ELT \citep{kasper10} which will
provide imaging, spectroscopy and polarimetry with a systematic contrast of 10$^{-9}$ at 100 mas separation.

\begin{acknowledgements} 
This research has benefitted from the M, L, T, and Y dwarf compendium housed at DwarfArchives.org. Use was
also made of the L and T dwarf data archive of Sandy Leggett, the IRTF spectral library and the Database of Ultracool parallaxes
maintained by Trent Dupuy. I thank Daniel Cotton and Brett Addison for valuable comments on the manuscript.

The work is supported by the Australian Research Council through Discovery grant DP110103167.

\end{acknowledgements}


\end{document}